\documentstyle[epsf]{elsart}
\begin{document}
%
%
\newcommand{\dg}{$^{\circ}\,$}
\def \qinj {\mbox{${\rm Q_{inj}}$}}
\def \chidof {\mbox{$ \chi^2/$DoF\ }}

\newcommand{\nue}{\mbox{$\nu_{e}\,$}}
\newcommand{\nuebar}{\mbox{$\overline{\nu}_{e}\,$}}
\newcommand{\numu}{\mbox{$\nu_{\mu}\,$}}
\newcommand{\numubar}{\mbox{$\overline{\nu}_{\mu}\,$}}
\newcommand{\nutau}{\mbox{$\nu_{\tau}\,$}}
\newcommand{\Mnue}  {\mbox{${\rm M}_{\nu_{e}}\,$}}
\newcommand{\Mnumu} {\mbox{${\rm M}_{\nu_{\mu}}\,$}}
\newcommand{\Mnutau}{\mbox{${\rm M}_{\nu_{\tau}}\,$}}
\newcommand{\Mnux}{\mbox{${\rm M}_{\nu_{x}}\,$}}
\newcommand{\nubar}        {\mbox{$\overline{\nu}$}}
\newcommand{\nux}          {\mbox{$\nu_x$}}
\newcommand{\nuxbar}       {\mbox{$\overline{\nu}_x$}}
\newcommand{\nul}          {\mbox{$\nu_l$}}
\newcommand{\nulbar}       {\mbox{$\overline{\nu}_l$}}
\newcommand{\nuone}        {\mbox{$\nu_1$}}
\newcommand{\nutwo}        {\mbox{$\nu_2$}}
\newcommand{\nuthree}      {\mbox{$\nu_3$}}

\newcommand{\ctau} {\mbox{$\langle c\tau \rangle\,$}}

\newcommand{\dEdx} {\mbox{${\rm d}E/{\rm d}x \,$}}

\newcommand{\MeV}   {\mbox{${\rm MeV}\,$}}
\newcommand{\GeV}   {\mbox{${\rm GeV}\,$}}
\newcommand{\MeVc}  {\mbox{${\rm MeV}\,$}}
\newcommand{\GeVc}  {\mbox{${\rm GeV}\,$}}
\newcommand{\MeVct} {\mbox{${\rm MeV}\,$}}
\newcommand{\GeVct} {\mbox{${\rm GeV}\,$}}
\newcommand{\pbarn} {\mbox{${\rm pb}^{-1}$}}
\newcommand{\fbarn} {\mbox{${\rm fb}^{-1}$}}

\newcommand{\ms} {\mbox{${\rm ms}$}}
\newcommand{\us} {\mbox{$\mu{\rm s}$}}
\newcommand{\ns} {\mbox{${\rm ns}$}}
\newcommand{\microns} {\mbox{$\mu{\rm m}\,$}}

\newcommand{\Fnu}          {\mbox{$\Phi(\nu)$}}
\newcommand{\Fnul}         {\mbox{$\Phi(\nul)$}}
\newcommand{\Fnux}         {\mbox{$\Phi(\nux)$}}
\newcommand{\Fnue}         {\mbox{$\Phi(\nue)$}}
\newcommand{\Fnuebar}      {\mbox{$\Phi(\nuebar)$}}
\newcommand{\Fnuenuebar}   {\mbox{$\Phi(\nue+\nuebar)$}}
\newcommand{\Fnumu}        {\mbox{$\Phi(\numu)$}}
\newcommand{\Fnumubar}     {\mbox{$\Phi(\numubar)$}}
\newcommand{\Fnumunumubar} {\mbox{$\Phi(\numu+\numubar)$}}
\newcommand{\Fpip}         {\mbox{$\Phi(\pi^+)$}}
\newcommand{\Fpim}         {\mbox{$\Phi(\pi^-)$}}
\newcommand{\Fmup}         {\mbox{$\Phi(\mu^+)$}}
\newcommand{\Fmum}         {\mbox{$\Phi(\mu^-)$}}

\newcommand{\Enu}          {\mbox{$E_{\nu}$}}
\newcommand{\dmsq}         {\mbox{$\delta m^2$}}

\newcommand{\DO} {\mbox{${\rm D}_2{\rm O}$}}
\newcommand{\HO} {\mbox{${\rm H}_2{\rm O}$}}
\newcommand{\Cl} {\mbox{$^{37}{\rm Cl}$}}
\newcommand{\Ga} {\mbox{$^{71}{\rm Ga}$}}
\newcommand{\B}  {\mbox{$^8{\rm B}$}}
\newcommand{\Ch} {\v{C}erenkov~}
%
%
\newcommand{\ra}{\rightarrow}
\newcommand{\la}{\leftarrow}
\newcommand{\da}{\downarrow}
\newcommand{\mlc}{\multicolumn}
\newcommand{\uln}{\underline}
\newcommand{\bi}{\begin{itemize}}
\newcommand{\be}{\begin{enumerate}}
\newcommand{\ei}{\end{itemize}}
\newcommand{\ee}{\end{enumerate}}
%
%
\newcommand{\eg}{{\it e.g.}}
%
%
\newcommand{\simge}{\mathrel{%
   \rlap{\raise 0.511ex \hbox{$>$}}{\lower 0.511ex \hbox{$\sim$}}}}
\newcommand{\simle}{\mathrel{
   \rlap{\raise 0.511ex \hbox{$<$}}{\lower 0.511ex \hbox{$\sim$}}}}
%
%
%
%
\def\mypub#1#2{#1, #2}
\def\mypre#1#2#3{#1\newline #2, #3}
%
%
%
%
\def\Journal#1&#2&#3(#4){#1 {\bf #2}, #3 (#4)}
%
%
%
%
%
%
\def\NIM{Nuclear Instruments and Methods }
\def\NIMA{Nuclear Instruments and Methods A }
\def\NPB{Nuclear Physics B }
\def\PLB{Physics Letters B }
\def\PRL{Physical Review Letters }
\def\PRD{Physical Review D }

\def\etal{{\it et al.}}

\begin{frontmatter}
   \title{The Sudbury Neutrino Observatory}
   \collab{The SNO Collaboration}

   %
%


   \author{J. Boger,}
   \author{R.L. Hahn,}
   \author{J.K. Rowley}
    
   \address{Chemistry Department, Brookhaven National Laboratory,  Upton, NY
            11973-5000 USA\thanksref{funding_DOE}}

   \thanks[funding_DOE]{Supported by the US Department of Energy.}


  \author{A.L. Carter,}
  \author{B. Hollebone,}
  \author{D. Kessler}

  \address{Carleton University,
           Ottawa, Ontario K1S 5B6 CANADA\thanksref{funding_NSERC}}

  \thanks[funding_NSERC]{Supported by Natural Sciences and Engineering Research
                         Council of Canada.}


   \author{I. Blevis,}
   \author{F. Dalnoki-Veress,}
   \author{A. DeKok,}
   \author{J. Farine\thanksref{Farine},}
   \author{D.R. Grant,}
   \author{C.K. Hargrove,}
   \author{G. Laberge,}
   \author{I. Levine,}
   \author{K. McFarlane,}
   \author{H. Mes,}
   \author{A.T. Noble,}
   \author{V.M. Novikov,}
   \author{M. O'Neill,}
   \author{M. Shatkay,}
   \author{C. Shewchuk,}
   \author{D. Sinclair}

  \address{Centre for Research in Particle Physics, Herzberg Laboratory,
           Carleton University,
           Ottawa, Ontario K1S 5B6 CANADA\thanksref{funding_NSERC}}

  \thanks[Farine]{Supported by the Swiss National Science Foundation in 1997.}


% ...

\vspace*{\fill}

   \begin{abstract}


The Sudbury Neutrino Observatory is a second generation water
\Ch detector designed to determine whether the currently observed
solar neutrino deficit is a result of neutrino oscillations.  The
detector is unique in its use of $\DO$ as a detection medium,
permitting it to make a solar model--independent test of the neutrino
oscillation hypothesis by comparison of the charged- and
neutral-current interaction rates.  In this paper the physical
properties, construction, and preliminary operation of the Sudbury
Neutrino Observatory are described.  Data and predicted operating
parameters are provided whenever possible.


   \end{abstract}


\end{frontmatter}


\section{Introduction}
\label{sec:introduction}

\indent The Sudbury Neutrino Observatory (SNO) has been constructed to
study the fundamental properties of neutrinos, in particular the mass
and mixing parameters.  Neutrino oscillations between the
electron-flavor neutrino, $\nue$, and another neutrino flavor have
been proposed~\cite{GribovPontecorvo} as an explanation of the
observed shortfall in the flux of solar $\nue$ reaching the earth, as
compared with theoretical expectations~\cite{BahcallDavis}.  SNO can
test that hypothesis by measuring the flux of $\nue$ which are
produced in the sun, and comparing it to the flux of all active
flavors of solar neutrinos detected on earth in an appropriate energy
interval.  Observation of neutrino flavor transformation through this
comparison would be compelling evidence of neutrino mass. Non-zero
neutrino mass is evidence for physics beyond the Standard Model of
fundamental particle interactions~\cite{StandardModel}.

The long distance to the sun makes the search for neutrino mass
sensitive to much smaller mass splittings than can be studied with
terrestrial sources.  Vacuum oscillations can change the ratio of
neutral-current to charged-current interactions, produce spectral
distortions, and introduce time dependence in the measured rates.
Furthermore, the matter density in the sun is sufficiently large to
enhance the effects of small mixing between electron neutrinos and mu
or tau neutrinos.  This matter, or MSW~\cite{MSW}, effect also applies
when solar neutrinos traverse the earth, and may cause distinctive
time and spectral modulations of the signal of $\nue$ in the SNO
detector. Measurement of these effects, made possible by the high
counting rate and separable $\nue$--specific and flavor--independent
responses of the SNO detector, will permit the determination of unique
mass and mixing parameters when combined with the existing results of
the light-water
\v{C}erenkov detectors, $\Cl$, and $\Ga$ solar neutrino
experiments~\cite{Homestake,Kamiokande,SAGE-GALLEX}.

The SNO experiment is unique in that it utilizes heavy water, $\DO$,
in a spherical volume of one kilotonne as a target.  The heavy water
permits detection of neutrinos through the reactions
\begin{eqnarray}
   & \nu_x  + e^-  \rightarrow & \nu_x + e^-    \label{eq:ES} \\
   & \nue   + d    \rightarrow & e^-   + p + p  \label{eq:CC} \\
   & \nu_x  + d    \rightarrow & \nu_x + n + p  \label{eq:NC}
\end{eqnarray}
where $\nu_x$ refers to any active flavor of neutrino.  Each of these
interactions is detected when one or more electrons produce \Ch light
that impinges on a phototube array.  The elastic scattering (ES) of
electrons by neutrinos (Eq.~\ref{eq:ES}) is highly directional, and
establishes the sun as the source of the detected neutrinos. The
charged-current (CC) absorption of $\nue$ on deuterons
(Eq.~\ref{eq:CC}) produces an electron with an energy highly
correlated with that of the neutrino.  This reaction is sensitive to
the energy spectrum of $\nue$ and hence to deviations from the parent
spectrum. The neutral-current (NC) disintegration of the deuteron by
neutrinos (Eq.~\ref{eq:NC}) is independent of neutrino flavor and has
a threshold of 2.2 MeV.  To be detected, the resulting neutron must be
absorbed, giving a $6.25$~MeV photon for absorption on deuterium or
photons totalling 8.6~MeV for absorption on $^{35}$Cl with MgCl$_2$
added to the $\DO$.  The photon subsequently Compton scatters,
imparting enough energy to electrons to create \Ch light.  (Once the
special--purpose neutral-current detectors have been installed, they
will provide the primary neutron detection mechanism.  See
Section~\ref{sec:NCDs} for details.)  Measurement of the rate of the
NC reaction determines the total flux of $\B$ neutrinos, even if their
flavor has been transformed to another active flavor (but not if to a
sterile neutrino). The ability to measure the CC and NC reactions
separately is unique to SNO and makes the interpretation of the
results of the experiment independent of theoretical astrophysics
calculations.

All experiments performed to date have detected fewer solar neutrinos
than are expected from standard solar
models\cite{introduction:SSM}. There are strong hints that this is the
result of neutrino flavor transformation between the production point
in the sun and the terrestrial detection point.  However, these
conclusions are based on reference to a calculated prediction.
Through direct measurements of the $\nue$--specific CC reaction and
the flavor--independent NC reaction, the SNO detector will be the
first experiment to make a solar--model--independent measurement of
the solar neutrino flux.

SNO can make contributions, some of which are
unique, in other areas of physics.  An example of the latter is a
search for the relic supernova neutrinos integrated over all past
supernovae.  For the relic supernova neutrinos, the interaction of
$\nuebar$ on protons in the deuterons produces a  \v{C}erenkov signal plus
two  neutrons, which makes a clean signature. Other topics that
SNO will study include the flavor composition of the
atmospheric neutrino flux, searches for certain types of dark matter, and
nucleon decay.  The observatory is designed to handle
the high data rates induced by an intense burst of neutrinos from a supernova
explosion as close as one kiloparsec away.

\begin{figure}[tbh]
   \setlength{\epsfxsize}{0.6\textheight}
   \centerline{\epsffile{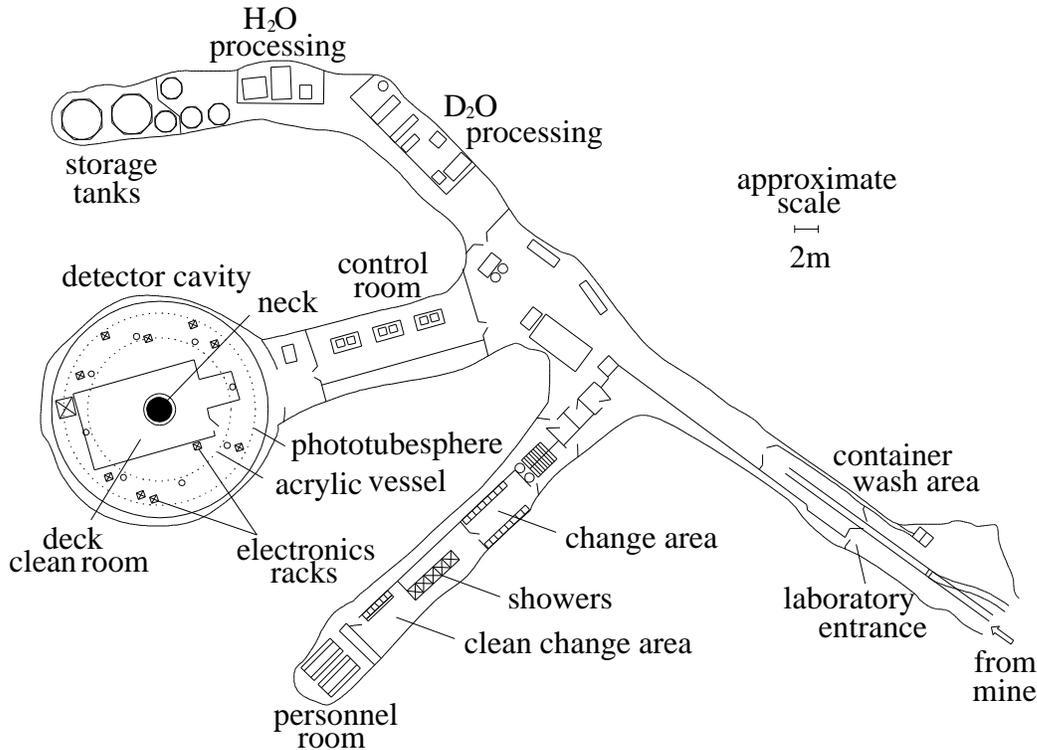}}
   \caption{General Layout of the SNO Laboratory.}
   \label{fig:LabLayout}
\end{figure}

\begin{figure}[tbh]
   \setlength{\epsfxsize}{0.8\textheight}
   \centerline{\epsffile{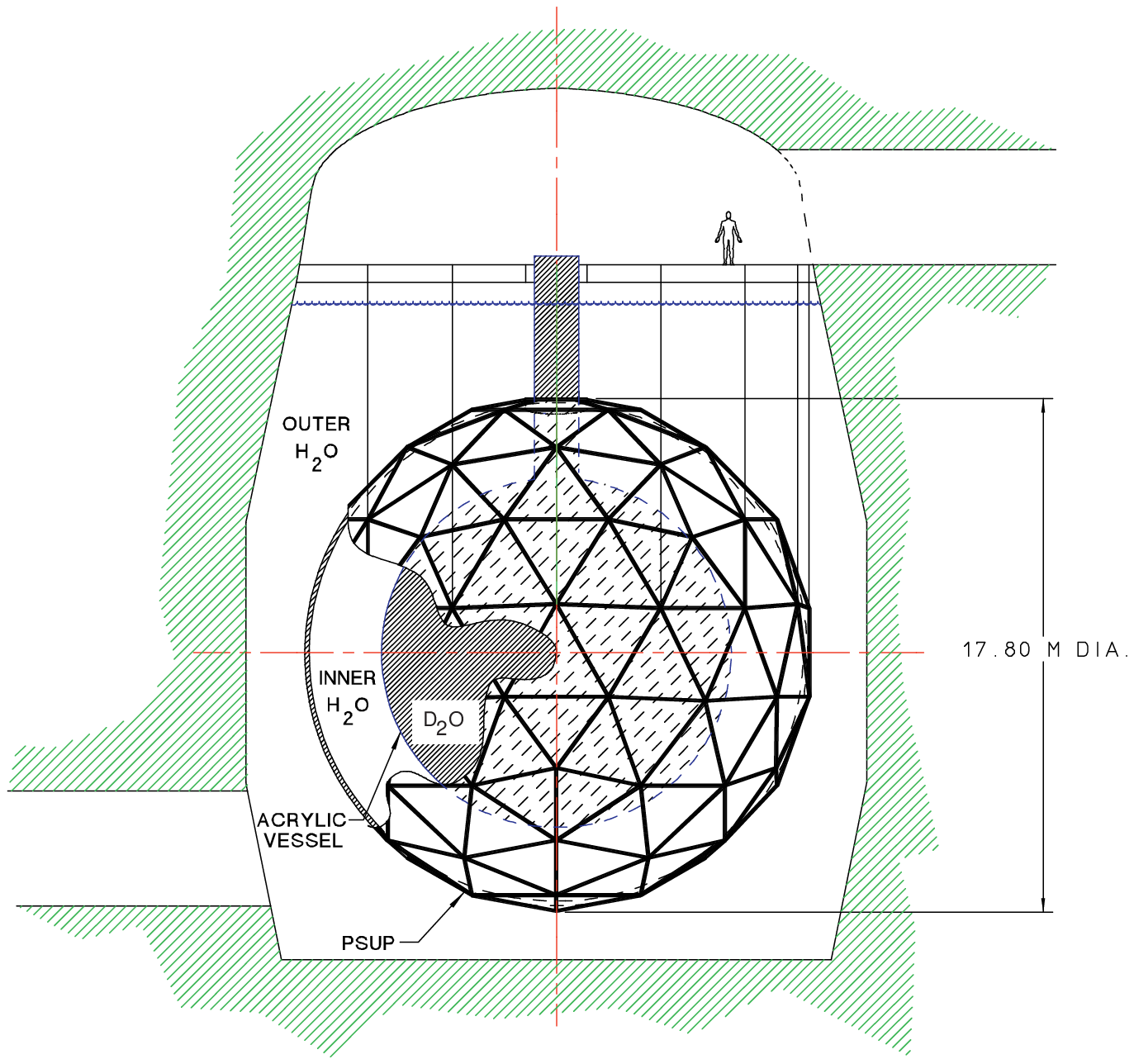}}
   \caption{The PMT support structure (PSUP) shown inside the SNO
	   cavity, surrounding the acrylic vessel, with light water
	   and heavy water volumes located as indicated.}
   \label{fig:PSUP}
\end{figure}

An overview of the SNO laboratory and detector is given in
Fig.\ref{fig:LabLayout}.  The detector consists of a transparent
acrylic sphere 12~m in diameter, supported from a deck structure by ten
rope loops made of synthetic fiber, as shown in Fig.~\ref{fig:PSUP}.
The sphere holds 1000~tonnes of heavy water.  Surrounding the acrylic
vessel is a geodesic structure 17.8~m in diameter made of
stainless-steel struts and carrying 9438 inward-looking
photomultiplier tubes.  The barrel-shaped cavity, 22~m in diameter and
34~m in height, is filled with purified light water to provide support
and shielding.

The SNO detector is located at $46^\circ 28' 30''$~N, $81^\circ 12'
04''$~W in the INCO, Ltd., Creighton mine near Sudbury, Ontario,
Canada.  The center of the detector vessel is 2039~m below the surface
in a cavity excavated from the ``6800-foot'' level of the mine.  The
surface is 309~m above sea level.  The granitic rock overburden
(mainly norite) corresponds to approximately 6000~m water
equivalent. At this depth, only about 70 muons pass through the
detector (inside the photomultiplier array) per day.

The sections below detail the design criteria of many of the important
aspects of the detector.  The extensive $\DO$ and $\HO$ purification
systems are described in Section~\ref{sec:water-systems} and 
purity specifications are given there.  Since SNO will add high purity MgCl$_2$
to the $\DO$ to enhance the NC detection, a description of the system
required to handle the heavy water brine is also given.  The
technically challenging construction of the large acrylic vessel (``AV'') which
holds the $\DO$ is described in Section~\ref{sec:AV}.  The large array
of PMTs and their basic operating parameters are described in
Section~\ref{sec:PMTs}.  The materials and construction of the PMT
support structure are described in Section~\ref{sec:PSUP}.  

A description of the electronic readout chain is given in
Section~\ref{sec:electronics}, followed by a description of the
downstream data acquisition hardware and software in
Section~\ref{sec:DAQ}.  Separate $^3$He neutral-current detectors,
which will be installed at a later date in the $\DO$ volume, are
described in Section~\ref{sec:NCDs}.  The detector control and
monitoring system is described in Section~\ref{sec:CMA}.  The
calibration system, consisting of a variety of sources, a manipulator,
controlling software, and the analysis of calibration data, is
described in Section~\ref{sec:calibration}.  The extensive effort to
maintain and monitor site cleanliness during and after construction is
described in Section~\ref{sec:cleanliness}.  The large software
package used to generate simulated data and analyze acquired data is
described in Section~\ref{sec:SNOMAN}.  Brief descriptions of the
present detector status and future plans are in
Section~\ref{sec:Conclusion}.

\section{Water Systems}
\label{sec:water-systems}

The SNO water system is comprised of two separate systems: one for the
ultrapure light water ($\HO$) and one for the heavy water
($\DO$). These systems are located underground near the detector.  The
source of water for the $\HO$ system is a surface purification plant
that produces potable water for the mine. Underground, the water is
pretreated, purified and degassed to levels acceptable for the SNO
detector, regassed with pure N$_2$, and finally cooled before it is
put into the detector.  Ultrapure water leaches out soluble components
when in contact with solid surfaces. It may also support biological
activity on such surfaces or on suspended particles.  The $\HO$ is
therefore deoxygenated and continuously circulated to remove ions,
organics and suspended solids. Both liquids are also assayed
continuously to monitor radioactive contaminants.

\begin{figure}[t]
   \setlength{\epsfxsize}{0.6\textheight}
   \centerline{\epsffile{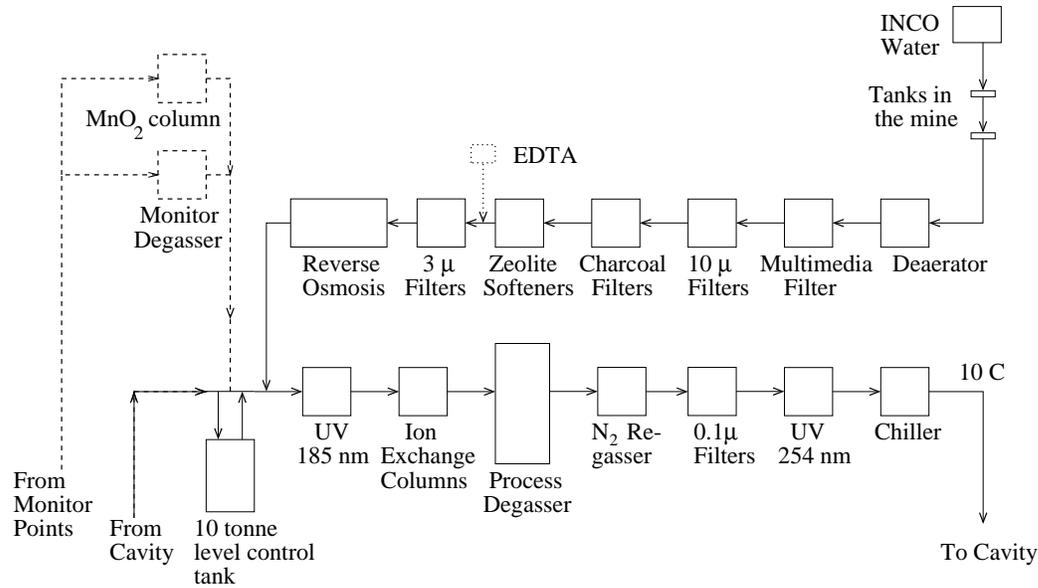}}
   \caption{The light water system.}
  \label{fig:water_H_system}
\end{figure}

Incoming potable water contains sand and silt particles, bacteria,
algae, inorganic salts, organic molecules and gasses (N$_2$, O$_2$,
CO$_2$, Rn, etc.).  After falling a total of 6800 feet, the water is
supersaturated with air, so first it enters a deaerator tank (see
fig.~\ref{fig:water_H_system}) where it spends a few minutes so that
some of the dissolved O$_2$ and N$_2$ comes out.  It then passes into
a multimedia filter consisting of a bed of sand and charcoal to remove
large particles followed by a 10-micron filter to remove fine
particles.  The water then enters  the laboratory water utility
room.  A charcoal filter is used to reduce the levels of organic
contaminants and to convert free chlorine into chloride since chlorine
would damage the reverse osmosis (RO) unit further downstream.

   After the charcoal filter, the water passes into 
softeners consisting of two 0.14~m$^3$ bottles containing strong-base
Purolite C100-E cation exchange resin. Here divalent ions such as Ca and Mg
are exchanged for Na ions.  The softeners also remove iron and
stabilize colloidal particles so they do not coagulate when
concentrated by the RO membranes.

A 9.1\% solution of sodium ethylene diamine tetraacetate (EDTA) and sodium
bisulphate is injected  at 9~ml/min. to complex various ion species (e.g., Al)
and to reduce O and Cl into a form that can be rejected by the RO.  Then two
filter units, each containing twelve 25-cm long, 3-$\mu$m filters, remove
suspended particles.  A silt density index test is done at this point
on the running system.

The reverse-osmosis process is the workhorse of the purification
system. Twelve spiral-wound thin film composite (polyamide on
polysulfone) membranes each 5.6~m$^2$ in area reduce inorganic salt
levels by a factor of at least 20 and reduce organics and hence the
EDTA and particles larger than molecular weight 200 with greater than
99\% efficiency. The RO performance is monitored online by percentage
rejection (typically 97.5\%) and conductivity monitors.  After the
detector is full the RO does not have to be used again unless SNO
requires substantial amounts of make-up water in the detector.

After the RO, the water enters a 185-nm UV unit consisting of mercury
lamps and quartz sleeves where any remaining organic compounds are
broken apart into ionic form.  The water next goes to an ion-exchange
unit that removes remaining dissolved ionized impurities left by the
RO.  These are two sets of six bottles in parallel containing
0.1~m$^3$ of Purolite nuclear grade NWR-37 mixed (cation and anion)
bed resins. The exiting water has a resistivity of 18.2~M$\Omega$-cm.

A custom--designed Process Degasser (PD) is used to reduce the O$_2$
and Rn levels by factors of about 1000 and 50, respectively, in the
water~\cite{water:PD}. The PD consists of a large electropolished
stainless steel vessel (81~cm diameter by 6~m high) containing shower
heads, spherical polypropylene packing and heater elements and pumped
with a mechanical booster pump (Edwards EH500A) backed by a four-stage
positive displacement rotary pump (Edwards QDP80 Drystar).  Vacuum is
maintained at 20~torr and water vapour flow rate at one~kg/hr.  The PD
removes all gases from the water and can cause, by diffusion, low
pressures inside the underwater PMT connectors.  Low pressure
compromises the breakdown voltage of the connectors, and in a
successful effort to reduce breakdowns that occurred with degassed
water, the water was regassed with pure nitrogen to atmospheric
pressure at the 2000-m depth using a gas permeable membrane unit.
This unit is followed by 0.1-$\mu$m filters to remove
particulates. Then a 254-nm UV unit is used to kill bacteria. Finally
a chiller cools the water to 10$^\circ$C before water is put into the
detector at a rate of 150~l/min.

The water mass between the AV and the PMTs is about 1700 tonnes and the mass
between the PMTs and the cavity walls is 5700 tonnes.  Water enters
the detector between the AV and PMTs. Because this region has to be
cleaner than the region outside, a 99.99\% leak-tight plastic barrier
seals the back of the PMTs.  Water in the outer region is dirtier due
to its large content of submersed material (cables, steel support,
cavity liner, etc.).  Water is drawn from this region back to the
utility room and into a recirculation loop.  This loop consists of a
ten-tonne polypropylene tank (used by the control loop to maintain
constant water level in the detector), the first UV unit, the ion
exchange columns, the process degasser, the N$_2$ regassing unit, the
0.1-$\mu$m filters, the second UV unit and the chiller.  The
recirculated light water is assayed regularly for pH, conductivity,
turbidity, anions, cations, suspended solids, dissolved gases, and
radioactivity.  This is accomplished by means of six sample pipes
in the $\HO$ volume.

\begin{figure}[t]
   \setlength{\epsfxsize}{0.6\textheight}
   \centerline{\epsffile{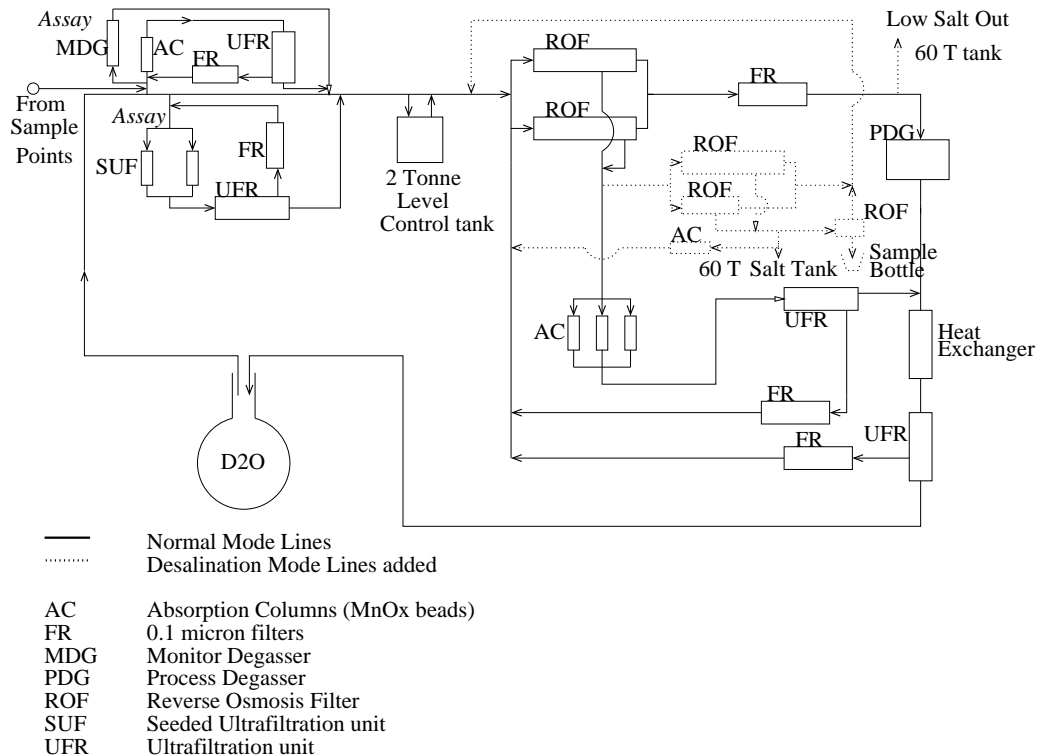}}
   \caption{The heavy water system.}
  \label{fig:water_D_system}
\end{figure}

  The heavy water system is designed to perform the following functions:
\begin{itemize}

   \item Receive the $\DO$ and make an initial purification to reduce
         the amount of contamination reaching the main system;

   \item Purify the $\DO$ with or without the MgCl$_2$ additive; 

   \item Assay the $\DO$ or $\DO$ brine to make an accurate background
         determination;

   \item Manage the addition and removal of the MgCl$_2$
         additive on a time scale short relative to the expected
         running times;

\end{itemize}
A diagram of the system is shown in Fig.~\ref{fig:water_D_system}.

The source of $\DO$ was the Ontario Hydro Bruce heavy water plant
beside Lake Huron.  The $\DO$ was trucked to the SNO site and
transported underground.  It was cleaned and stored temporarily before
it was put into the acrylic vessel (AV).  Heavy water delivered to the
lab was first passed through ion-exchange columns to reduce its ionic
content, in particular its K content.  The $\DO$ then went into a
large polypropylene-lined tank.  During the filling of the $\HO$ and
$\DO$, the $\DO$ was filled at such a rate as to maintain zero
pressure differential across the bottom of the AV, which assured that
the vessel surface was generally in compression.

  With the SNO detector filled, the $\DO$ is recirculated to
maintain its purity. This is accomplished by an RO
system consisting of five separate pressure housing units and
the membranes contained within. Two large (22~cm diameter and 5.5~m
length) units are used in parallel for the purification of the
recirculating heavy water.  The concentrate stream containing the
radioactive ions passes through adsorption columns to decrease the
Th, Ra and Pb concentrations.

All the water returning to the AV passes through an ultrafiltration
unit.  This is particularly important following MnO$_{\rm x}$ and
seeded ultrafiltration assays (see below) to insure that fines from
these absorbers are not carried into the AV.  The $\DO$ is assayed
regularly for density, pH, conductivity, turbidity, anions, cations,
total organic carbon and suspended solids.
Table~\ref{table:water_isotope} gives the isotopic composition of the
heavy water.

\begin{table}
   \caption{Isotopic composition of SNO heavy water.}
   \label{table:water_isotope}
   \medskip
   \begin{tabular}{|l|c|l|c|} \hline
      Isotope & Abundance & Isotope & Abundance \\ \hline
      $^{2}$H & 99.917(5)\% &  $^{17}$O & 0.17(2)\%  \\
      $^{3}$H &  0.097(10) $\mu$Ci/kg &  $^{18}$O & 0.71(7)\% \\
      $^{1}$H & Balance &  $^{16}$O & Balance \\ \hline
   \end{tabular}
\end{table}

The $\DO$ and $\HO$ have to be isolated from the laboratory air to
prevent radon (at a concentration of 3 pCi/l) from getting in, even in
the presence of large pressure transients that occur during mine
operations.  A cover gas system provides nitrogen gas that functions
as a physical barrier between the water and the radon-rich laboratory
air. The nitrogen comes from the boil-off of liquid nitrogen stored in
a 1000-liter dewar. The boiloff gas has been measured to contain less
than $10^{-5}$~pCi radon per liter of nitrogen gas.

Radon in the lab can also enter the water system in other ways.
Therefore, when a filter is changed on the $\DO$ system, radon-free
$\DO$ already in the $\DO$ system will be used to flush the filter to
get rid of trapped laboratory air before bringing the filter back on
line.  In the $\HO$ system, $\HO$ losses from the process degasser,
filter changes and small leaks (if any) in the Urylon cavity liner is
made up with radon-free water that has been ``aged'' in a storage
tank.

The radioactivity requirements are set primarily by the design goal
specifying that the contribution to the neutron-production rate
through photodisintegration of the deuteron, with an interaction
threshold of 2.2~MeV, should not exceed one-tenth that of the model
solar-neutrino rate (5000 y$^{-1}$). This leads to combined limits of
$3\times10^{-15}$g/g of Th and $4.5\times10^{-14}$g/g U in the
$\DO$. SNO's assay systems are expected to have an uncertainty of not
more than 20\% in their measurement at these impurity levels, although
plating may give rise to a systematic uncertainty. There are six
sample pipes in various locations inside the AV that allow $\DO$
samples to be withdrawn for radioassay.

For Th, Ra and Pb assay the seeded ultrafiltration~\cite{water:SUF}
technique (SUF) will be used.  The SUF unit contains four large microfilters
thinly coated with DTiO.  One hundred tonnes of water at a rate of 150
l/min  will be passed through them.  Then the filters are removed and
the Th, Ra, and Pb stripped off and counted.  Acrylic beads coated
with MnO$_{\rm x}$ have been shown to be an excellent adsorber for Ra
with an extraction efficiency of about 90\%~\cite{water:RaAdsorber}.
Water is passed at 20~l/min. through a 1-l column containing the
MnO$_{\rm x}$ which is removed off-line for counting of the Ra daughters
using an electrostatic device in which charged ions are deposited on
the surface of a silicon detector~\cite{water:silicon_detector}.

The number of $^{222}$Rn (3.8-d half-life) is measured in six
tonnes of liquid taken from one of six regions in the $\HO$ or one of
six regions in the $\DO$.  The six tonnes of water are put through a
small vacuum degasser~\cite{water:degasser} at 20 l/min.  The gasses
which are removed from the water are primarily N$_2$, O$_2$, Ar,
CO$_2$ and a few hundred atoms of $^{222}$Rn, as well as 10 ml/min in the form 
of water vapour.  Rn is subsequently frozen out in a U-shaped trap cooled
with liquid nitrogen (-192 C) and transferred to a ZnS coated
scintillation cell~\cite{water:ZnS} for counting.

To enhance the neutral-current detection, MgCl$_2$ will be added to
the water to take advantage of the larger neutron-capture cross
section of Cl relative to deuterium.  A concentration of approximately
0.2\% will be used.  A $\DO$ brine solution that has been purified
prior to transport underground is put into a large polypropylene lined
tank. When it is time to add the MgCl$_2$ to the AV, this tank will
slowly be emptied. At the same time, an equivalent volume of salt-free
$\DO$ will be taken out of the AV and put into a second polypropylene
lined tank.

To take the MgCl$_2$ out of the AV, the third RO unit will be used in
combination with the two main ones to desalinate the water to about
100~parts per million (ppm). In a second desalination pass, a fourth,
smaller RO unit will be added to the process to desalinate the $\DO$
to about one~ppm.

\section{Acrylic Vessel}
\label{sec:AV}

\begin{figure}[t]
   \setlength{\epsfxsize}{0.7\textheight}
   \centerline{\epsffile{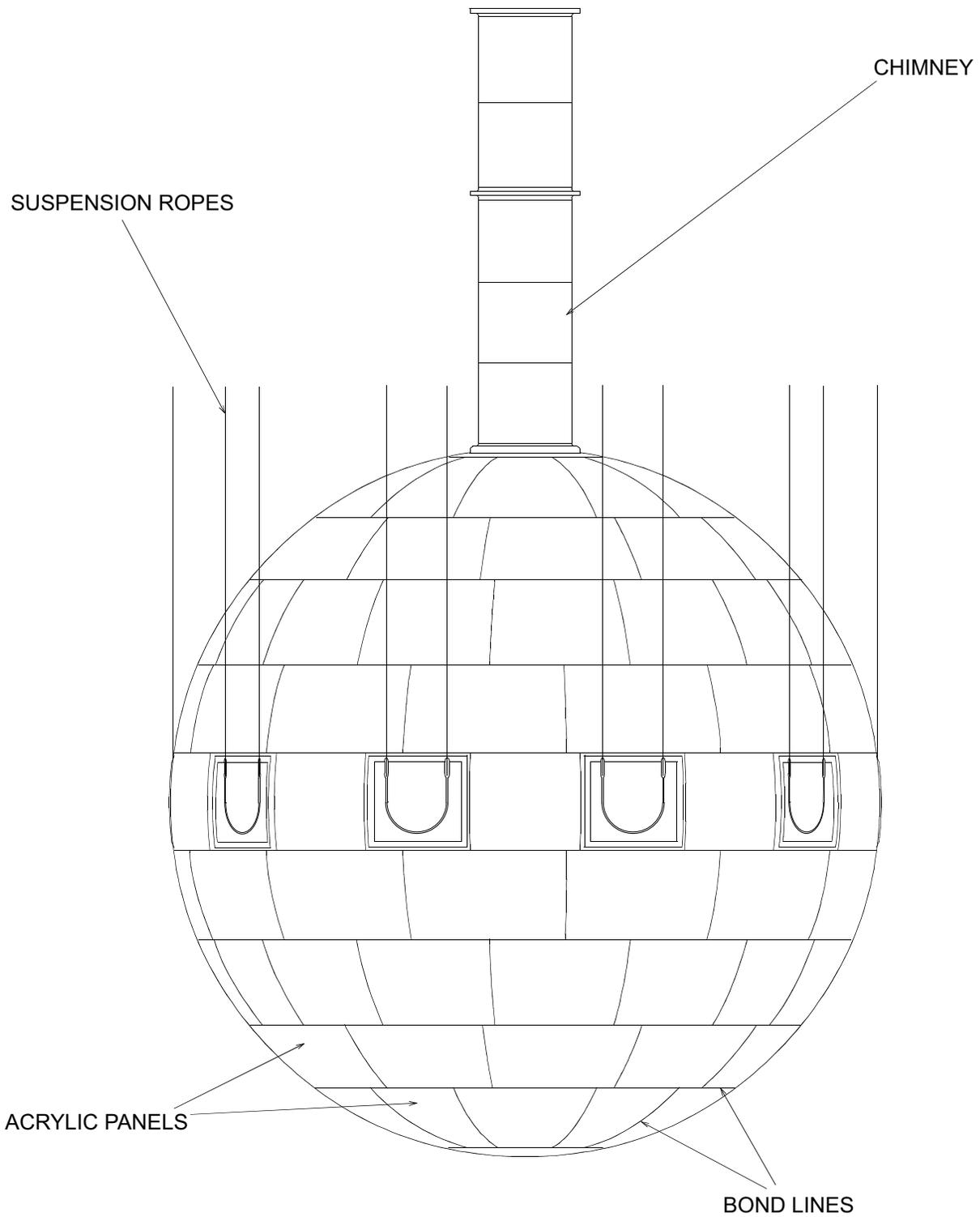}}
   \caption{The SNO acrylic containment vessel.}
   \label{fig:AV}
\end{figure}

The $\DO$ containment vessel must meet diverse
requirements, some of which present opposing design constraints. The
primary design criteria for the containment vessel are:
\begin{itemize}
\item  Isolate 1000 tonnes of $\DO$ from surrounding $\HO$.
\item  Maintain structural integrity and performance over  ten years while
immersed in ultrapure $\DO$ and $\HO$ and
   subjected to the seismic activity expected in an operating mine.
\item  Minimize the total mass
   of radioactive impurities.
\item  Maximize optical performance. 
\item  Design for construction in the mine.
\end{itemize}
The design criteria listed above resulted in the containment vessel
shown in Fig.~\ref{fig:AV}. 

A 12 meter diameter sphere was chosen as the optimum shape for the
containment vessel. A sphere has the largest  volume-to-surface ratio, and
 optimum stress distribution. This reduces the mass of acrylic needed, and
hence the amount of radioactivity.
      The sphere is suspended from ten loops of rope, which are attached to
the vessel by means of rope grooves located around the equator of the
sphere. The choice of rope suspension was driven by radioactivity and
optical considerations: a rope under tension offers excellent load
bearing capacity for a minimum mass, and it presents the minimum
obstruction of the \v{C}erenkov light that neutrino interactions produce.

To allow the insertion of calibration devices and the installation of
the $^3$He neutron detector strings, the sphere is provided with a
1.5-m diameter by 6.8-m tall chimney.  Piping also enters and
exits the vessel through this chimney.  For filling and purification
recirculation requirements  7.6-cm diameter pipes introduce $\DO$ at
the bottom of the vessel and remove it from the top of the
chimney. $\DO$ may be extracted via 3.8-cm diameter pipes from four
different levels in the vessel and one in the chimney for
measurements of both chemical and radioactive water quality.

A total of 122 ultraviolet transmitting (UVT) acrylic panels were used
in the construction of the spherical part of the containment
vessel. These panels were nominally 5.6~cm thick, with the exception
of the ten equatorial panels containing grooves for the suspension
ropes. These rope groove panels were nominaly 11.4~cm thick.  Acrylic
sheet was chosen as the construction material for a number of reasons.
A simple hydrocarbon, UVT acrylic can be manufactured with very low
intrinsic radioactivity.  The light transmission of UVT acrylic
matches reasonably the spectral response of the PMTs.  Cast acrylic
sheet is commercially available in sizes acceptable for mine
transportation. It is readily thermoformed into spherical segments and
is easily machined. It is capable of being bonded together with bond
strengths close to that of the parent material.

The flat cast acrylic panels were first thermoformed into spherical
sections by slump forming into a female mold formed from polished
aluminum plate with a radius of 6.06~m (the outer radius of the
sphere).  The panels were then machined to the correct shape on a
five-axis milling machine while at a constant temperature of
21$^\circ$C.  To avoid contamination, only clean water was used as a
lubricant during machining operations.

A final check of the dimensions was made by ``dry assembling'' the
panels on a special framework to form first the upper hemisphere, then
the lower hemisphere. This ensured that all the panels could be fitted
together within the specified tolerance of $\pm$25~mm on the
spherical curvature required by the buckling criteria.  During
construction the curvature was typically maintained to $\pm$6~mm.

The panels were bonded together using a partially polymerized adhesive
that cures at room temperature.  This adhesive was formulated by
Reynolds Polymer Technology Inc., the fabricators of the acrylic
vessel. The bonds are approximately 3~mm thick and special allowance
has to be made for the 20\% shrinkage during polymerization of the
adhesive.  It took considerable R\&D and construction time to obtain
over 500~m of adequate quality bonds.  All bond strengths exceeded
27.5~MPa. Prior to bonding the main vessel the bonding techniques were
prototyped by building two small (seven-panel) sections of the sphere.

First the upper hemisphere of the vessel was built and the chimney
attached. The chimney of the containment vessel was made of five cylinders of
ultraviolet absorbing (UVA) cast acrylic to reduce the ``piping'' of
unwanted \v{C}erenkov light from these regions into the inner volume
of the detector.

The hemisphere was then raised into its final position and
suspended on its ten supporting ropes. Vectran fibers were chosen as the
material for the suspension ropes, not only for their low radioactivity but
also for their ability to retain strength during long term exposure to
ultrapure water. A total of 300~kg of Vectran filament was supplied by Hoechst-Celanese
Company.  The filaments were twisted into rope by Yale Cordage. Due to
the large surface area of the filaments this represents a significant
contamination potential due to dust deposition. Therefore the twisting
machines were carefully cleaned and tented in plastic prior to use.
 Each of the ten loops supporting the vessel is approximately
30~m long and 24.4~mm in diameter.  In normal operation, each loop is
continuously loaded to five tonnes, or approximately 10\% of its ultimate
strength of 500,000~N.

Working from a suspended construction platform, subsequent rings of
the lower hemisphere were constructed and attached to the hanging
upper hemisphere. As construction progressed the platform was lowered
until the final ``south pole'' disk was installed.

The principal features of the containment vessel are
listed in Table~\ref{table:AV_features}.
\begin{table}[tb]
      \caption{Principal features of the containment vessel (all numbers are at 20$^\circ$C).}
      \label{table:AV_features}
      \begin{tabular}{|l|l|} \hline
        Capacity of sphere           & 997.91~tonnes of $\DO$ \\
        Vessel internal diameter     & 12.01~m \\
        Nominal wall thickness       & 5.5~cm \\
        Mass of sphere               & 30.0~tonnes \\
        Capacity of chimney (normal operation)     & 8.61~tonnes of $\DO$ \\
        Chimney internal diameter    & 1.46~m \\
        Chimney height               & 6.8~m \\
        Mass of chimney              & 2.53~tonnes \\
        Bulk absorption coeff. of light in acrylic & .01~cm$^{-1}$ @400~nm \\
        					    & .04~cm$^{-1}$ @360~nm \\
        					    & .18~cm$^{-1}$ @320~nm
\\\hline
      \end{tabular}
\end{table}

If acrylic is subject to excessive stress for extended periods of time
it will develop crazing cracks which eventually will lead to premature
failure. The ten-year design life of the vessel requires that the
long-term tensile stresses not exceed 4~MPa. In order to reduce
further the tensile stresses, it was decided to place the vessel in
compression during normal operation by adjusting the $\HO$ level with
respect to the $\DO$ level.  Optimization of the design was carried
out with the ANSYS~\cite{av:ANSYS} finite-element analysis code.  The
structure was studied under a variety of simulated conditions, both
normal and abnormal (e.g., with a broken suspension rope). For all
these studies it was assumed that the mechanical properties of the
bonds were identical to that of the acrylic, an assumption supported
by measurements of bonded test specimens.

The Polycast Corp. measured mechanical properties of the acrylic
panels to insure that each sheet met design specifications for
thickness and mechanical integrity.  The averages of five measurements
are listed in Table~\ref{table:AV_mechanical}.  The low percentage of
residual monomer listed in the table indicates that the polymerization
process is essentially complete and that the mechanical properties of
the acrylic will not change.
\begin{table}[tb]
      \caption{Mechanical measurements of the acrylic panels.}
      \label{table:AV_mechanical}
      \begin{tabular}{|l|l|} \hline
        Tensile strength        &        79~MPa      \\
        Tensile elongation      &        5.05\%      \\
        Tensile modulus         &        3.5~GPa     \\
        Compressive deformation &        0.36\%      \\
        Residual monomer        &        1.22\%     \\ \hline
      \end{tabular}
\end{table}

Stress levels in bonded panels were also recorded by positioning crossed
polarisers on each side of the bond and photographing fringes. At the end, the
vessel was successfully proof-tested by reducing the internal pressure  7~kPa
below ambient pressure to subject the sphere to buckling forces and by
pressurizing it to 14~kPa above ambient pressure to subject all the bonds to
tensile stresses.

The radioactive and optical requirements for the acrylic were
established by Monte Carlo simulation of the detector. To ensure that
the acrylic would meet the requirements of the detector, the
radioactive properties were measured throughout the supply of
materials by means of neutron activation, mass spectrometry, and alpha
spectroscopy.  Concentrations of Th and U in each acrylic sample were
measured to be less than the specified 1.1~pg/g each. For the
eleven ropes produced, twelve 2.5-kg samples were taken from the
beginning and end of a production run and between each rope.  Direct
gamma counting yielded upper limits of 200~pg/g Th and U, in agreement
with the average value of small-sample neutron-activation results.

Optical absorption coefficients for each production batch of acrylic
sheets were measured for samples cut from the sheets. Over 300
measurements were performed.  A useful figure of merit is the ratio of
the light detected with acrylic to the light detected without acrylic
between 300~nm and 440~nm at normal incidence.  The figure of merit
for the ten thicker equatorial rope groove panels was calculated from
the absorption coefficient as if they were 5.6~cm thick.  For the 170
sheets manufactured the average figure of merit is 0.73.

\section{Photomultiplier Tubes}
\label{sec:PMTs}

The primary design considerations for the PMT system are
\begin{itemize}
   \item  High photon detection efficiency,

   \item  Minimal amount of radioactivities in all components, 
      $< 120$~ng/g  U, $< 90$~ng/g Th, $< 0.2$~mg/g K,

   \item  Low failure rate for a 10-year lifespan submerged in ultrapure water at a pressure
      of 200~kPa and for the seismic activity expected at the SNO
      site,

   \item  Fast anode pulse rise time and fall time and low photoelectron
      transit time spread, for a single-photoelectron timing
      resolution standard deviation $< 1.70$~ns,

   \item  Low  dark current noise rate, $< 8$~kHz, at a charge gain of $10^{7}$,

   \item  Operating voltage less than 3000~V,

   \item  Reasonable charge resolution, $> 1.25$ peak to valley,

   \item  Low  prepulse, late-pulse and after-pulse fractions, $< 1.5\%$,

   \item  Low sensitivity of PMT parameters to external magnetic field: at 100~mG, 
      less than 10\% gain reduction and less than 20\% timing
      resolution degradation.

\end{itemize}
Raw materials from the manufacturers of the PMT components, bases,
cables, and housings were assayed for radioactivity~\cite{pmt:guelph}.
The leach rates of different types of glass and plastic were also
measured.

The photomultiplier tubes (PMTs) are immersed in ultrapure water to a
maximum depth of 22~m, corresponding to a maximum water pressure of
200~kPa above atmospheric pressure. The failure rate of the PMT
components, which consist of the PMT, the resistor chain and the
HV/signal cable, must be compatible with the expected approximate ten
year detector lifetime. This puts severe contraints on the choice of
glass for the PMT envelope, the shape of the envelope, the type of
cable, and the design of the waterproof enclosure that protects the
resistor chain.  Raw materials from all manufacturers were assayed for
radioactivity~\cite{pmt:guelph}.  The dominant contributor to the
contaminant content of the PMT is the glass which has a mass of 850~g
and Th and U impurity levels of roughly 40~ppb.  The leach rates of
different types of glass and plastic were also measured.

Following detailed stress analysis of the proposed glass envelopes, a
mushroom shaped envelope was judged to be most suitable for this
underwater application and least likely to deteriorate under long term
hydraulic pressure in UPW.  Schott Glaswerke produced a new borosilicate glass
(Schott 8246) for SNO. The Th and U impurities in this glass are less
than 40~ppb. Its water resistance is rated as class 1 (same as
Pyrex). In addition, this glass has outstanding optical transmission
properties at short wavelengths, with a bulk light absorption
coefficient of better than $0.5$~cm$^{-1}$ at 320~nm wavelength, and
low He permeability.  Schott refitted a furnace  with a special
low-radioactivity liner to produce 16000 mouthblown bulbs for
SNO and the LSND Project\cite{pmt:LSND}.

The most important PMT parameters are the noise rate, the efficiency,
the transit-time spread and the amount of K, U and Th in each PMT. The
energy resolution and the event vertex spatial resolution are largely
determined by the first three parameters and the detector energy
threshold is strongly affected by the radioactivities in the PMT
components.

\begin{figure}[t]
   \vspace{8cm}
   \includegraphics{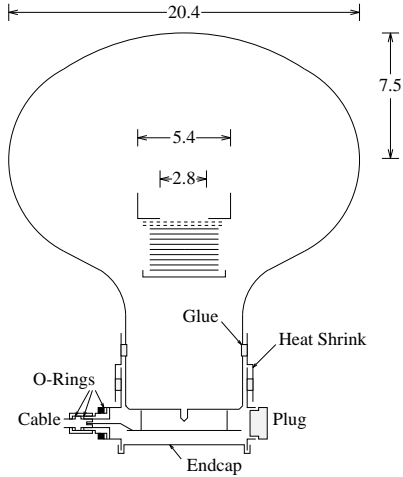}
   \caption{A schematic of the Hamamatsu R1408 PMT and its waterproof
      enclosure. The 9 dynodes are shown as solid horizontal
      lines. The focusing grid is shown as the dashed lines above the
      dynodes. The volume inside the enclosure is filled with soft
      silicone gel. The end cap is flexible to allow for thermal
      expansion. The plug is sealed with heat shrink and thermal
      adhesive.  Dimensions are in cm.}
  \label{fig:PMT}
\end{figure}

The Hamamatsu R1408 PMT was selected for use in SNO.  A schematic
drawing of the R1408 is shown in Fig.~\ref{fig:PMT}.  From the
measured U and Th concentrations in the internal parts and glass
envelope, the total weight of Th and U in each PMT is estimated to be
about 100 ${\mu}$g, about a factor of 14 below specifications. The
9438 inward facing PMTs collect the \v{C}erenkov photons, providing a
photocathode coverage of 31\%. To improve the light collection
efficiency, a 27~cm diameter light concentrator is mounted on each PMT,
increasing the effective photocathode coverage to about 59\%.  The
reflectivity of the concentrator reduces this figure to 54\%.  Of the
inward facing tubes, 49 have a dynode tap and a second signal cable.
These ``low gain'' channels extend the dynamic range of SNO at high
light intensities by about a factor of 100.  Another 91 PMTs without
concentrators are mounted facing outward to detect light from muons
and other sources in the region exterior to the PMT support structure.

The PMT anode is at positive high voltage, typically in the range
1700~V to 2100~V, and the photocathode is at ground voltage.  A single
RG59/U type cable is used to carry the high voltage and the fast anode
pulse which is capacitively coupled to the front-end electronics. The
primary concern in selecting materials for the resistor chain circuit
board is radioactivity.  Each two-layer Kapton circuit board, with
surface mounted components and a through-hole 0.0047-$\mu$F film
capacitors for filtering and source termination, contains a total of
about 2~${\mu}$g of thorium and 0.6~${\mu}$g of uranium, about a
factor of five lower than the design goal.

The main function of the waterproof enclosure is to keep the PMT base
dry when the PMT is deployed under water. There are two water barriers
protecting the circuit board to ensure a low failure rate. The outer
enclosure consists of a plastic housing, waterproof modified ``TNC''
connector and heat shrink tubing with thermal adhesive. (A connector
was used to simplify construction and cabling.) The shrink tubing and
adhesive are used to hold the plastic housing onto the PMT. The cavity
inside the enclosure is filled with a silicone dielectric gel (GE
RTV6196) which acts as the second water seal.  The silicone gel
contains less than 23 ng/g of Th and less than 12 ng/g of U. The
polypropylene plastic housing contains less than 10~ppb of Th and less
than 6~ppb of U. The heat shrink tubing is made of clear polyolefin
which has low Th and U impurities.  The TNC connector was provided by
M/A-COM with modifications specified by SNO.  These connectors
exhibited breakdown at nominal high voltage settings, a problem which
was overcome by the addition of nitrogen gas to the water, as
described in the previous section.  Heat shrink and thermal adhesive
seal the cable to the male connector.

The RG59/U type cable was designed by Belden, Inc. for
SNO. It consists of a copper clad steel central conductor surrounded
in succession by 1.85~mm thick solid polyethylene dielectric, 95\%
tinned copper braid, Duofoil bonded metallic shield and 1.3~mm thick
high-density polyethylene outer jacket. The colouring compound mix in
the outer jacket is reduced to 0.1\% carbon black, the minimum amount needed to
make the jacket opaque. Each PMT cable is 32~m long and contains 
between 10--17~${\mu}$g of Th and a few~${\mu}$g of U.  The attenuation at
400~MHz is 7.3~dB$/$30.5~m. The delay is 4.9~ns$/$m.

The Hamamatsu R1408 PMT has unusually stable gain characteristics with
respect to the influence of weak magnetic fields~\cite{pmt:ra}. In the
SNO detector the earth's magnetic field is about 55 ${\mu}$T and
points approximately 15 degrees off the vertical axis. In such a
field, the photon-detection efficiency average over all the PMTs would
be about 82\% of the detection efficiency at zero magnetic
field. Because the PMT efficiency is reduced by less than 3\% at
magnetic field intensities of 20 ${\mu}$T, 
only the vertical component of the magnetic field in the SNO detector
is cancelled with 14 horizontal field-compensation coils embedded in the cavity
walls. With the proper currents, the maximum residual field in the PMT
region is 19 ${\mu}$T, and the reduction in photon-detection
efficiency is about 2.5\% from the zero-field value.

The single photoelectron test system and test results are described in
detail in refs.~\cite{pmt:jillings,pmt:chris}. A total of 9829 PMTs
passed the acceptance test.  The average RMS timing resolution is
found to be 1.7~ns, in line with the specification.  The mean
relative efficiency, defined in Ref.~\cite{pmt:jillings}, is 10\%
better than specification. The mean noise rate is 2.3~kHz at
$20^\circ$C. The cooler temperature in the detector (approximately
$10^\circ$C) reduces the mean noise rate to approximately 500~Hz,
including signals due to residual radioactivity in the detector. The
mean operating voltage is 1875~V.

\section{Photomultiplier Tube Support Structure}
\label{sec:PSUP}

The photomultiplier tubes, their light concentrators and associated
hardware are collectively referred to as the PMT Support Structure or
the PSUP.  The geometry of this platform is fundamentally established
by maximizing photon collection while minimizing background signals,
fabrication costs, complexity, and transportation and installation
intricacy.  The $\DO$ target geometry, the PMT specifications, light
concentrator performance, and cavity geometry all affect the final
PSUP configuration.

The PSUP serves the additional function of providing a barrier between
the core of the experiment (the $\DO$ target and light collection
surfaces) and the outer regions of the experiment.  This barrier
shields the PMTs from light generated in the outer regions of the
experiment.  These regions include the cavity walls and the support
piping and cabling for the experiment.  Near the cavity walls the
radioactivity in the surrounding rock creates a significant photon
background.  The complexity of the cabling and piping makes the outer
region difficult to clean and keep clean during construction and the
radioactive purity of the materials needed for some of these functions
could not be reasonably controlled to the same levels as other
detector components.  The PSUP also functions as a highly impermeable
barrier to waterborne contamination, ensuring that the highly purified
water in the sensitive region between the PMTs and the AV is
effectively isolated from the dirtier water in the outer regions.

The PSUP also supports  the
outward looking PMTs, LED calibration sources, and water-recirculation and
monitoring piping.

The design criteria for the PSUP include:
\bi
   \item Maximizing the collection of optical signals from the $\DO$ 
         target while minimizing a 
         variety of background sources;
   \item Maintaining performance and integrity of the array,
         with no required maintenance,
         for at least a ten year span
         submerged in ultrapure water and
         for a variety of seismic conditions
         anticipated at the SNO site;
   \item Minimizing the mass of the components to reduce 
         intrinsic backgrounds;
   \item Producing and fabricating the array from documented low 
         radioactivity materials;
   \item Maintaining a high impermeability water barrier between the inner and 
         outer light water regions;
   \item Maintaining an effective barrier to light generated in the 
         outer regions;
   \item Minimizing the complications of transporting and installing the 
         array at the underground  site;
   \item Producing and fabricating the array from materials inimical to biological growth with low 
         leaching characteristics, low 
         magnetic susceptibility and 
         low electrolytic characteristics;
   \item Utilizing materials and installation processes compatible with the 
         underground (mining) environment;
   \item Installing the upper half of the geodesic sphere and 
         supporting the loads of 
         roughly half the detector array during which time the AV is constructed.
\ei

The PSUP is logically viewed as two systems--a geodesic sphere that
functions as the main support system and the panel arrays that house
the PMTs and concentrators.  The geodesic sphere, an 889-cm radius three-frequency
icosahedron, is shown in Fig.~\ref{fig:PSUP}.  Normally this geodesic sphere
would utilize 92 nodal connections between the 270 struts. This choice of
geodesic structure uses three different strut lengths with similar linear
dimensions. The topmost node is replaced with a hollow toroidal ring and the
connecting struts shortened to accommodate the acrylic vessel chimney. 
Additional toroidal rings guide the acrylic vessel support ropes through the
PSUP. Ninety-one outward viewing PMTs are supported on the PSUP struts.
The design of the geodesic sphere and panel arrays is fully
documented~\cite{psup:psupnotes}.

The geodesic sphere is suspended on 15 stainless steel wire rope
cables from the cavity deck.  These cables terminate in a spherical thrust
bearing on the deck and are monitored with load cells.  The position
of the array is adjustable in the z-direction (up-down) by $\pm$ 10~cm. When
submerged, the array becomes buoyant, and is therefore secured to the cavity
floor at a set of anchor points, each with a pulley and cable that attaches to
the main PSUP suspension brackets.

Wire rope cannot effectively be manufactured without lubrication.
The main suspension cables are plastic coated and encased in an
additional plastic housing to prevent  contamination of the water.  Anchor
cables are plastic coated.

\begin{figure}[t]
   \setlength{\epsfxsize}{0.6\textheight}
   \centerline{\epsffile{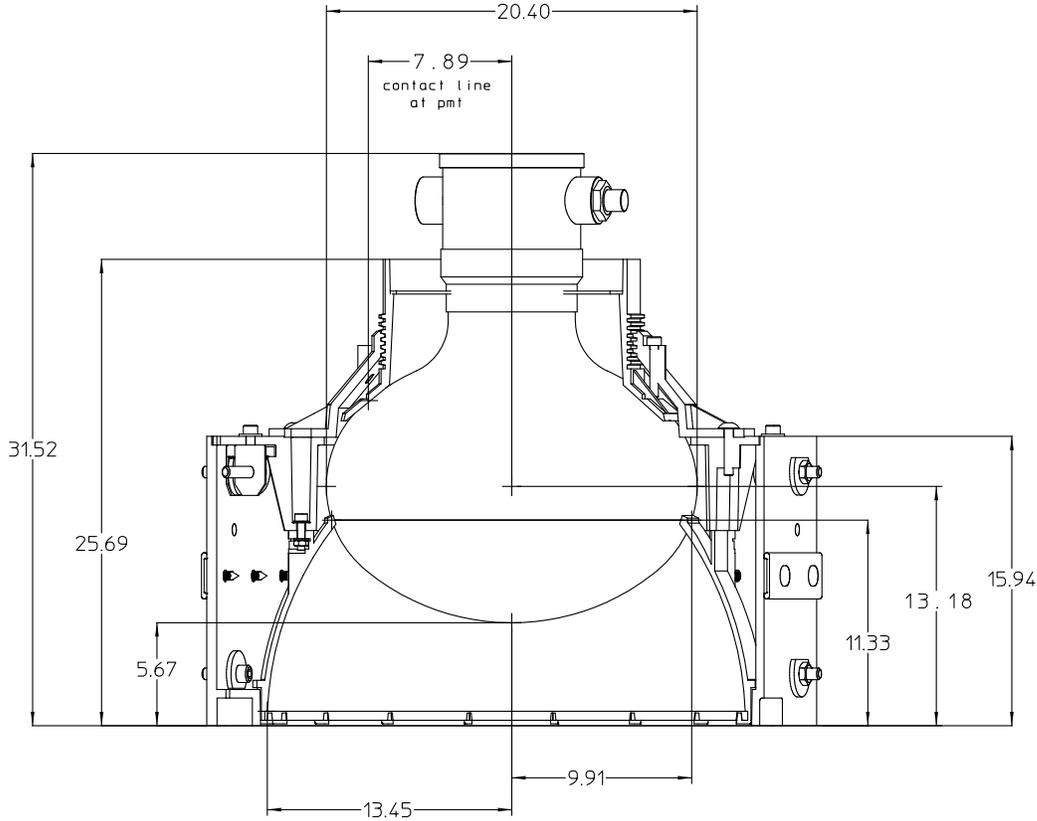}}
   \caption{Photomultiplier mounted in a hexagonal ABS cell, which also forms the
            light-concentrator housing. Dimensions are in cm.}  
   \label{fig:PSUPHexcell}
\end{figure}

The PMTs are secured in hexagonal ABS
(Acrylonitrile-Butadiene-Styrene) black plastic housings that also
support light concentrators (reflectors) for each inward-looking
PMT~\cite{psup:light_concentrators} as shown in
fig.~\ref{fig:PSUPHexcell}.

The concentrators increase the effective area of the photocathode to
maximize the number of photons detected, and limit the angular
acceptance of the photomultipliers so that only the central part of
the detector (where backgrounds are lowest) is in view. They are
designed with a three-dimensional critical angle of $56.4^\circ$,
which means the detector is viewed by each PMT out to a radius of
7~m. The increase in light collection efficiency is approximately
75\%~\cite{psup:light_concentrators}.

A concentrator is made of 18 pieces of thin dielectric-coated aluminum
sheet curved and placed around the holder of ABS plastic. The
optically active coating consists of a layer of specular aluminum
covered by a low refractive index layer of magnesium fluoride, which
is in turn covered by a high refractive index layer of mixed titanium
and praeseodymium oxides.  The coating thicknesses are optimized to
maximize reflectance in water over the visible and near-UV range of
wavelengths, and over a wide range of incident angles. The mean
effective reflectance in water is 82$\pm$3\%.

The top dielectric layer also forms a protective barrier against
corrosion in water and was applied to the back exposed surface of the
aluminum pieces as well.  A protective layer of titanium oxide was
applied to the edges. The material was subjected to extensive aging
prior to construction. It was found to have a useful life of many
years under deoxygenated deionized water with a pH around 7, the
normal environment in SNO.

The hexagonal PMT assemblies are collected together into flat arrays
of between 7 and 21 PMTs to form a set of repeating arrays used to
tessellate the spherical surface of the geodesic structure.  Each
panel array is secured to the geodesic structure with three adjustable
mounts.  These mounts permit the delicate panel arrays to accommodate
the geodesic structure struts flexing under changing load conditions
(most notably, when going from a fully instrumented dry array to a
fully submerged array).  This alignment is critical to ensure that
each PMT views the full $\DO$ volume.  A total of 751 panel arrays
tessellate the sphere providing 9522 hexagonal cells.  Of these
locations, the final design used 60 cells to accomodate the 20 AV
ropes and six to provide access at six locations for calibration
sources preserving the optical and water flow requirements for the
array. During the water fill 18 PMTs were removed for diagnostic
testing and their locations appropriately capped, leaving 9438 cells
occupied with PMTs.  The hexagonal cells cover $\sim85\%$ of the
surface area of the geodesic sphere.

Water piping and sampling tubes as well as several pulsed blue LED
calibration devices are
accommodated in the small spaces between PMT panel arrays.

The PMT support structure is subjected to a variety of loading conditions
and stresses.  Exposure to ultrapure water is known to reduce the
mechanical performance of many materials over time.  The design is
optimized with the criteria listed above with both hand calculations
and a detailed ANSYS finite element analysis of the PSUP under a
variety of conditions.  A half-scale model, partial full-scale model
and dry fit-up of the entire geodesic sphere were used to confirm the
design.

The analyses (both stress and buckling) study numerous effects, such
as the variation of mechanical and physical properties with time and
immersion in water, the effects of water fill on the deflections and
stresses in the PSUP in order to establish acceptable filling-rate
limits, the consequences of one of the PSUP supports failing, the
effects of various construction tolerances, and the response of the
natural frequency of the PSUP under dynamic
conditions~\cite{psup:standards}.

Various stainless steel alloys are used in the fabrication of the
geodesic sphere and panel arrays, primarily SS304L for welded
components and SS304 for parts that are only machined.  These alloys
have an excellent documented history of long-term exposure to
deionized (ultrapure) water. Welding was accomplished with inert-gas
processes. A vacuum-deposited Ag coat provides a clean anti-galling
agent for threaded fasteners without compromising water purity.

In order to reduce the likelihood of stress corrosion and cracking,
plate and tube material is pickled and passivated.  Steel fasteners
are fabricated from raw materials meeting the same specifications and
are in addition heat-treated and quenched.

ABS plastic is used in the injection molded components of the PMT
housings.  General Electric resin GPX5600 color 4500 was selected for
the molding.  This selection of plastic provides excellent impact,
thermal, chemical resistance, and injection molding characteristics.
Since ABS plastic has a limited documented history of exposure to
deionized water, we conducted extensive accelerated aging tests to
confirm the behavior of the ABS GPX5600 to long term ultrapure water
exposure and to insure an adequate safety factor for the design
life\cite{psup:psupnotes2}. Plastic components were injection molded
using controlled raw materials and processes.  The machines were
thoroughly cleaned prior to and during the fabrication cycle of the
components.  The assembly areas were tented and machine operators used
clean gloves during the production of the items.

Radioactivity of all components or samples of raw materials used in
the fabrication of the PSUP were subjected to direct counting to
determine U and Th concentration.  Specified limits of 15 ng/g for each of these
species are met by all materials; measured upper limits are typically a factor
of 3 to 5 times lower. All clean components were stored in clean
plastic bags prior to assembly. Assembly of the PSUP components was
accomplished in cleanroom assembly areas with conditions typically better than
Class 1000.   Assembled components were stored in clean plastic bags prior to
installation.

\section{Electronics}
\label{sec:electronics}

The electronics chain is required to provide sub-nanosecond time and
wide-dynamic-range charge measurement for the PMT pulses.  While the
solar neutrino event rate is very low, the electronics chain must
handle background rates in excess of 1~kHz and burst rates from
potential supernovae in excess of 1~MHz without significant deadtime.
The chain is implemented using three full custom, application specific
integrated circuits (ASICs) and commercial ADCs, memory, and
logic. The Data Acquisition (DAQ) interface is VME compatible.  The
ASICs carry out virtually all the analog signal processing and part of
the digital signal processing.  The ASIC chip set consists of a
wide-dynamic-range integrator, a fast and sensitive discriminator/gating
circuit, and an analog/digital pipelined memory with a timing circuit. The use
of these ASICs provides increased reliability and decreased per-channel cost. 
Further cost reduction and improved functionality is obtained by using
high-volume, large-scale-integration commercial components for the remainder of
the circuitry, including numerous field programmable gate arrays (FPGAs).

\begin{figure}[t]
   \setlength{\epsfxsize}{0.4\textheight}
   \centerline{\epsffile{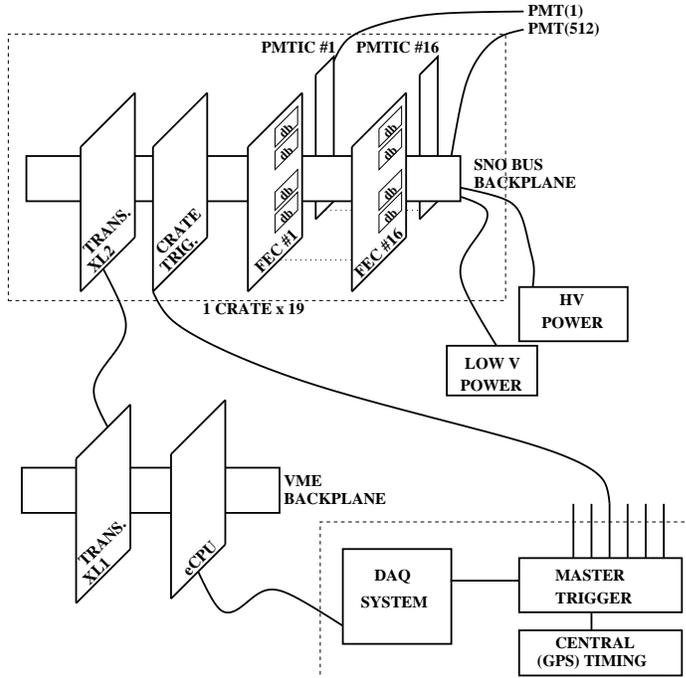}}
   \caption{Overview of the SNO custom data acquisition electronics.}
   \label{fig:electronics_overview}
\end{figure}

The signal processing electronics occupies 19 crates, each
processing signals from 512 PMTs.  The signal and high voltage for each PMT are
carried on a single cable and connected in groups of 8 at the rear of each crate
to one of sixteen PMT interface cards (PMTICs).  The signal then enters one of
sixteen front-end cards (FECs), where it is processed by the custom ASICs.  The
ASICs reside on four daughter boards (DBs), each of which handles 8 channels. 
One FEC processes 32 channels, digitizes the signals, and stores the digital
results in 4 MB of on--board memory, large enough to buffer event bursts
(roughly speaking, the memory in all the FECs could hold a one-million-event
supernova burst).  The FECs are attached to a custom backplane
(``SNOBus'') that implements the SNO signal- and power-distribution
protocol.  Also attached to the backplane are a translator card and a
trigger-formation card.  The XL1 translator card operates in conjunction
with a companion XL2 translator card in the DAQ VME crate to communicate
with the single-board embedded processor (eCPU) of the DAQ VME crate
downstream (see Section~\ref{sec:DAQ}).  The trigger card and the
associated trigger system are described below.
Programmable test,
calibration, and diagnostic facilities are implemented throughout the
crate.  An overview of the full system is provided by
Fig.~\ref{fig:electronics_overview}.

The analog part of the chain, from PMT to ASIC chip set, operates as
roughly 9500 independent, asynchronous, data-driven pipelines.  The
subsequent readout part of the chain operates as a separate,
clock-driven system that moves the data from the FECs to the DAQ VME
system.  Each element of the chain
is described below.

The analog pulse from each PMT travels through 32~m of 75$\Omega$
RG59-like waterproof coaxial cable to the PMTIC.
One commercial  high-voltage power supply provides about 70~mA DC for
the 16~PMTICs (512 PMTs) in one crate.
Current to the PMTs passes
through individual isolating, filtering, and trimming networks.  The
PMTIC plugs into the rear of the SNOBus crate and connects directly to
its companion FEC in the front of the crate.  The PMTIC also provides
disconnects for individual cables, HV blocking capacitors for the PMT
signal, overvoltage and breakdown protection for the integrated
circuits, limited readback of the PMT current, and a programmable
calibration pulse source for each channel. The PMT signal is
transmitted through a connector to the FEC where it is properly
terminated, split and attenuated.

\begin{figure}[bt]
   \centerline{\epsffile{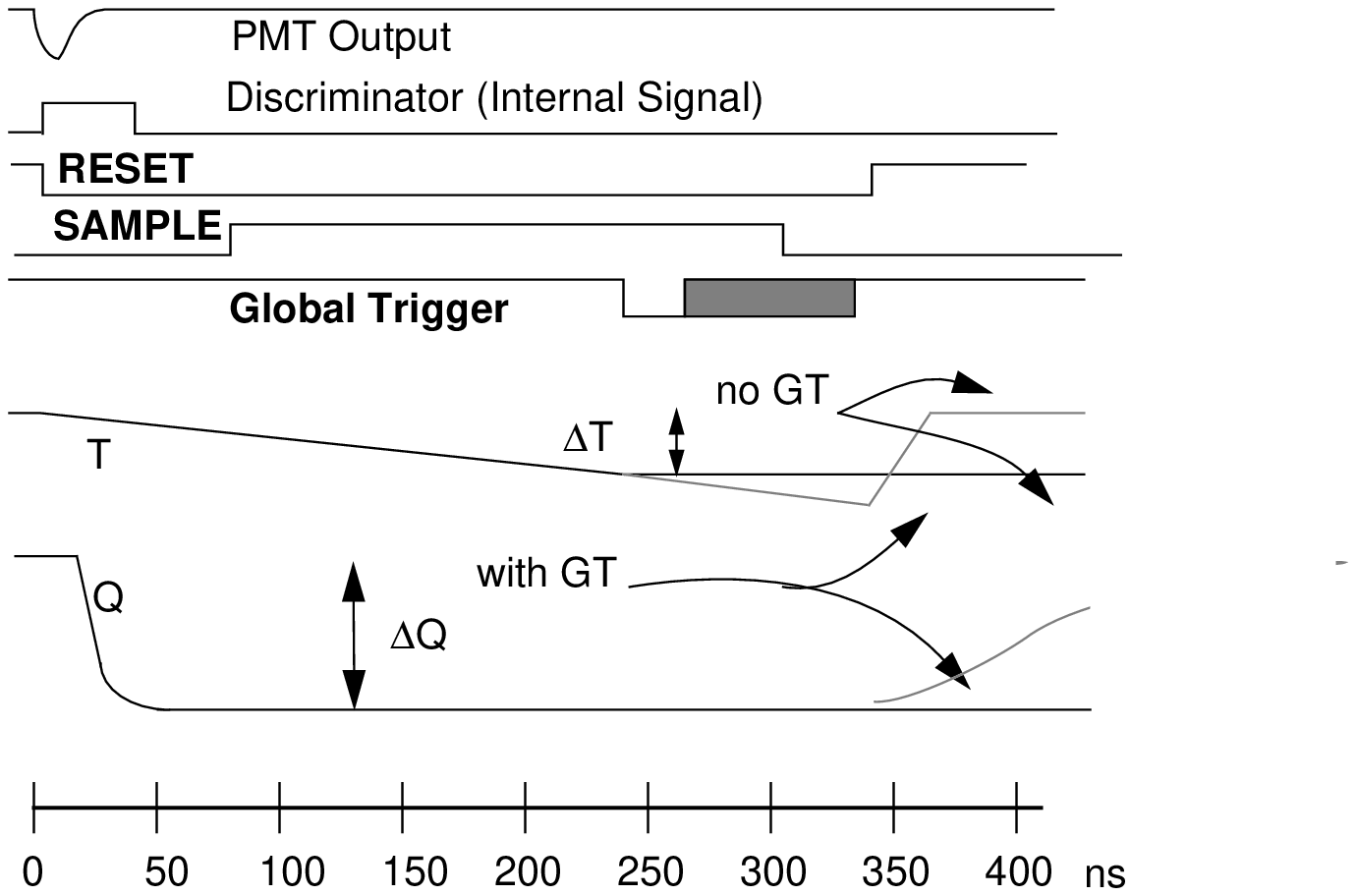}}
   \caption{Single channel timing cycle. With no GT present, a channel
            resets automatically at the end of a timing cycle ($\sim$ 400~ns).
            }
   \label{fig:electronics_timingdiag}
\end{figure}

On the FEC, the signal current is delivered to one of four daughter
boards.  The DBs physically separate most of the analog signals from
most of the digital signals in order to reduce potential crosstalk,
and simplify the overall FEC$+$DB design.  The signal is then fed to a
four-channel discriminator chip (SNOD) where any leading edge is
observed by a fast differentiator. The current is split into two
branches (approximately in the ratio 1:16) and fed into two separate
channels -- one low gain, the other high gain -- of an eight-channel
charge integrator (SNOINT).  The SNOD and SNOINT chips were fabricated
as custom designs in AT\&T's CBICU-2
process~\cite{electronics:CBICU,electronics:bipolar}. The SNOINT chip,
which uses high-quality external capacitors for the actual
integration, has dual integrator channels in order to increase the
effective dynamic range to more than 14 bits.  The SNOINT also
provides shaped low and high gain analog sums of the PMT input
signals for use in detector triggering.  Each channel of the SNOD chip
has independent discriminator and gate generators to provide the
timing functions necessary for the SNOINT chip.  A timing diagram for
the SNOD and SNOINT chips is shown in
Fig.~\ref{fig:electronics_timingdiag}.  The SNOINT and SNOD chip
outputs are then delivered to the CMOS ASIC analog memory.

The CMOS member of the SNO chip set (QUSN7) provides analog memory, a
time--to--amplitude converter (TAC), and channel and trigger logic for
the SNO detector.  The 16--deep memory  samples the low-  and high-gain
channels of the SNOINT at early and late times, for a total of four
possible charge samples~\cite{electronics:cmos}.  The QUSN7 chip also
receives a signal from the SNOD which is used to initiate a time
measurement cycle and the creation of system trigger primitives.
These single-channel trigger primitives constitute the first stage of
a 9438-input analog sum.
This mixed
analog/digital device uses custom analog blocks
combined with auto-placed and routed standard cells and is fabricated
in the Northern Telecom CMOS4S process~\cite{electronics:NT}.

The timing sequence and TAC in QUSN7 are initiated on the leading edge
of a signal from SNOD. The TAC for a given PMT is started whenever
that PMT fires, and is either stopped by a centrally generated Global
Trigger (GT) signal or resets itself if no GT arrives after an
internal timeout period.  On a valid GT, four analog voltages
(corresponding to one time and three of the four selectable charge
measurements) are stored in one of the 16 analog memory banks, an
associated digital memory records the sequence number of the GT (for
event building), the memory location (for second order corrections of
the data) and any associated condition flags, and a data-available
flag is set.  The overall trigger deadtime is less than 10~ns, and
since each channel is self-resetting, the only other relevant deadtime
is the {\em per PMT} deadtime, which is set to about 400~ns to allow
for light reflection across the SNO detector volume.  These two
contributions to the detector deadtime are negligible.

The QUSN7 chip also generates two trigger primitive signals, a short
coincidence (roughly 20 ns) and a long coincidence (roughly 100 ns),
some basic utility functions, and extensive self-test capability.
Separate internal counters are used to keep track of PMT noise rates
and count error conditions.  All counters and latches are accessible
for testing via an external scan path and a variety of programmable
adjustments are built in.

A data available flag from any QUSN7 initiates an external 32--channel
readout sequencer.  The sequencer is a clocked, synchronous state
machine implemented in a standard field-programmable gate array (FPGA).  At the
sequencer's convenience, a QUSN7 chip with available data is selected and
presents four individually buffered analog voltages to the four on-board 12-bit
2-$\mu$s ADCs.  After the ADCs have sampled the analog voltages, the
sequencer reads three bytes of digital information: the 16-bit GT
sequence number, four bits of cell address, and four flag bits.  A
fourth read strobe clears the memory location.

At the end of the sequencer's cycle, a 3-word, 12-byte, fixed-format data
structure is loaded into the on--board memory. The memory is a standard SIMM
DRAM under the control of a commercial dual port controller. The sequencer
operates one port of the controller as a FIFO while the second port is accessed
via the downstream DAQ interface. Each three word descriptor is complete in
the sense that the measured time and charge and the geographic channel
number and temporal GT sequence number are fully specified, so while it
is possible that hit descriptors will be loaded into memory out of
temporal sequence, they can be properly sorted out in the downstream
DAQ event builder.

The pair of translator cards (XL1, XL2) implements a high speed
($>8$~MB/s) RS485 link between each SNOBus crate and the central DAQ
VME crate (shown in Fig.~\ref{fig:electronics_overview} containing a
Motorola 68040 single-board computer, or ``embedded CPU'' (eCPU)).
The translator card located in the SNOBus crate performs a
TTL$\leftrightarrow$GTL conversion.  The use of GTL (Gunning
Transceiver Logic, with a 0.8-V swing) on the SNOBus backplane for all
data, address and control lines reduces the possibility of crosstalk
with the sensitive, low voltage PMT signals.  For the same reason,
clock, DTACK (the VME data handshake signal) and other important
signals are transmitted on the SNOBus backplane as low--level
single--ended emitter-coupled logic.  All signals are terminated, and
the use of a 2-mm grid, 165-pin connector on this custom backplane
makes for a more compact and flexible system and permits the use of a
greatly increased number of ground lines relative to standard VME.

Each MeV of energy deposited in the SNO detector is expected to result
in roughly eight contemporaneous photoelectrons detected by the PMT
array. Thus the most powerful hardware trigger for detection of solar
neutrinos is a simple count of the number of PMTs that have fired in
the recent past.  Because the PMT array is 17~m in diameter,
different photons from an interaction in the $\DO$ could differ in
travel time to the PMTs by as much as 66~ns (or longer, if they
undergo reflection(s)). For this reason the
trigger resolving time is set digitally to about 100~ns to allow all unreflected
photons from a \Ch event to be counted toward a possible trigger.  The
actual counting of hits is done via a chain of analog summations.

A signal from SNOD is used in QUSN7 to initiate a pair of independent
current pulses. The longer pulse (nominally 100~ns) is sent to the
first stage of a 9438-input analog sum.  Summations at the FEC,
crate, and full detector level follow.  The final summation is
compared against a programmable threshold, and a sum above this
threshold will generate a GT.  The second current pulse generated by QUSN7 is
nominally much narrower (about 20~ns), has a digitally set width and
delay and, in an identical summing tree, generates a
separate trigger useful for studies in which one wishes to scrutinize a
selected, smaller, fiducial volume of the detector.

In addition to these two discrete analog sums, the SNO trigger system
can trigger on the analog sum of the 9438~shaped PMT signals produced
by the SNOINT chips, and copies of each of the sums are sent to a
digital oscilloscope for real time monitoring of the state of all
9438~channels.  For the 100~ns coincidence trigger, up to three
separate thresholds can be set simultaneously, and triggers from the
lowest threshold can be prescaled with a factor as large as 65,534 to
provide a monitor of very low energy backgrounds.  A built-in pulser
provides a ``pulsed--GT'' trigger, which is used as a ``zero bias''
trigger for the detector.  All trigger types can be enabled
simultaneously, and bits corresponding to each type that fired in a
given event are latched in a ``trigger word'' that is part of the data
stream.

Two separate oscillators are used to keep a record of absolute and
relative time. A commercial GPS system provides a 10~MHz signal as
well as precision time markers at requested Universal Times.  This
system is nominally accurate to order 100~ns and will allow
correlation of SNO data with that of other astronomical detectors.
Since the GPS receiver must be on the surface and all the other
electronics is below ground, communication delays due to the 4 km long
fiber optics must be continuously monitored. This is done by using a
separate fiber in the same bundle and measuring total round trip
propagation time.  A highly stable quartz oscillator operating at 50
MHz offers a higher precision inter--event timing standard.

\begin{figure}[hbt]
   \vspace{11.5cm} 
   \includegraphics{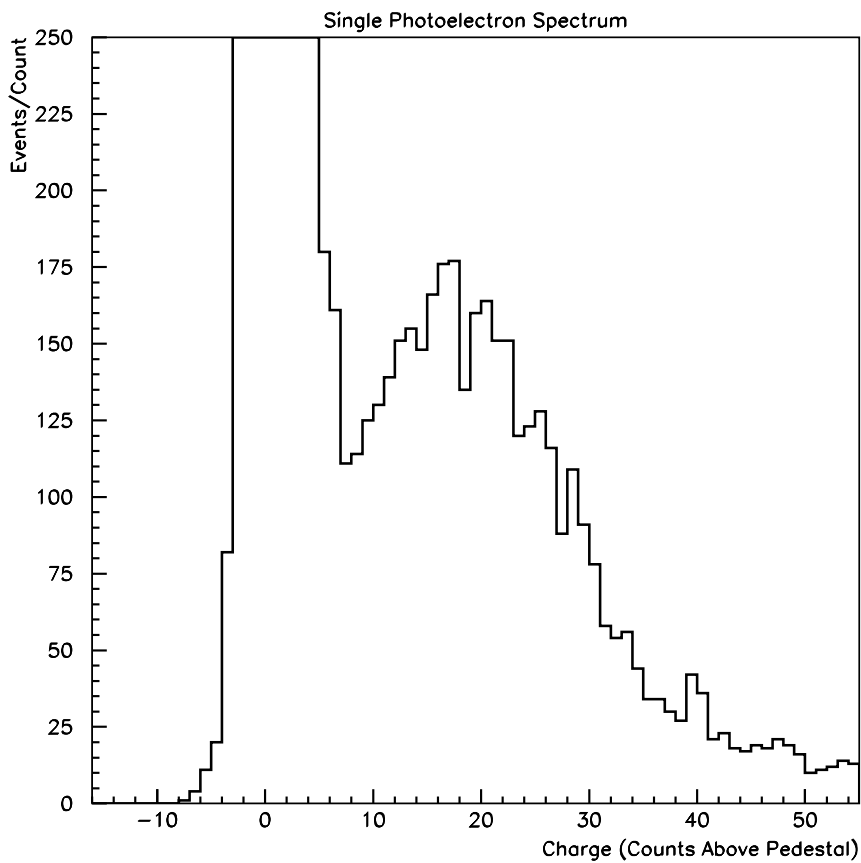}
   \caption{Single photoelectron spectrum using the full electronics
            chain with an {\it in situ} SNO PMT.  The truncated noise peak 
            reaches roughly 2000 counts.}
   \label{fig:electronics_SPE} 
\end{figure}

Extensive measurements have been made of the electronics all the way
from the integrated circuits to the full operating 9728 channel system
(19 crates with 512 channels each, a number of which are connected to
non-PMT sources) which had been installed underground and was
completely operational by the spring of 1998. Tests of the front end
custom ASICs running as single devices, as a coupled chip set, and on
full functionality production boards running with the prototype and
then production DAQ software were very successful. The performance of
the bipolar chips is very close to that predicted in the
pre-fabrication simulations and the CMOS chip is adequate for the
experiment. The charge linearity fully satisfies our design goals. The
linearity and precision of the time measurement are also significantly
better than the minimum requirements of SNO.  Tests with pulsers and
actual SNO PMTs have verified stable and reliable operation of the
chip set and the system even under the unanticipated challenge of high
rate full voltage breakdowns in degassed $\HO$ at some of the
underwater PMT connectors.  A single-photoelectron spectrum with no
discriminator threshold set is shown in
Fig.~\ref{fig:electronics_SPE}.

\section{Data Acquisition}
\label{sec:DAQ}

In this section the design and functioning of the SNO Data Acquistion
System (DAQ) is outlined. First, a general overview of the system
architecture is presented, followed by a more in-depth discussion of
the details of the readout and control, event-building, and low- and
high-level monitoring software.

The principal hardware components of the SNO DAQ system are a single
VME crate containing a Motorola 68040 single-board computer without an
operating system (the embedded CPU, or eCPU), but with 32~MByte of on-board
memory. A PCIbus-VME interface (the Macintosh Dual-Port Memory or
MDPM) with 8~MByte of memory, an SBus-VME interface  (the Sun
Dual-Port Memory or SDPM) with 8~MByte of memory, the Master Trigger
Card (MTC), and the seven XL1 cards needed to access the 19 ``SNO crates''
containing the front-end electronics also reside in this crate.

Each SNO crate is read out sequentially by the eCPU, which accesses
one crate in turn through the transparent XL1/XL2 interface bus.
Within each crate, the FECs are read out sequentially, by programming
the slot-select register on the crate's XL2 card.

Once the contents of the FEC memory have been read by the eCPU, they are
transferred to a circular buffer on the SDPM. The eCPU additionally
has access to the Master Trigger Card memory; once it has
checked every enabled FEC, it reads out the MTC memory and transfers
the trigger data to a separate circular buffer on the SDPM.

The SDPM is simultaneously and asynchronously accessed in read-only
mode by a Sun Ultra Sparc 1/170 workstation, which is responsible for
sorting the FEC and MTC data and building them into events. The Sun
writes full events to disk.  A separate mechanism is responsible for
copying completed event files to a tape drive located on this Sun as
well as copying, via Ethernet, these files to a Sun on the surface,
where they are copied again to tape.  The system also supports
monitoring programs running on a variety of platforms.  These monitoring
programs access a sampled data stream in real time over the network.

The readout of the electronics is both controlled and monitored via a
user interface Program, the ``SNO Hardware Acquisition and Readout
Control (SHARC),'' running on a Power Computing 250~MHz PPC Macintosh
compatible, which uses the MDPM to communicate with the eCPU. A set of small,
predefined data blocks transfer control and low-level monitoring information
such as user commands and eCPU status between SHARC and the eCPU.  SHARC 
initializes the hardware,  takes individual hardware elements (channels, cards,
crates) in and out of the readout,  stores the current hardware configuration
(which elements are present and which are being read out), and ensures that
this information is saved.

SHARC controls SNO's calibration sources through a
manipulator computer that moves the sources around inside the detector, and
also receives information regarding individual source status from the
various computers used to operate the different sources
themselves. The manipulator or ``calibration'' computer inserts this
slowly-varying data into the data stream by sending it to SHARC, which writes
it to a calibration data block in the MDPM. Source information that changes
more quickly (additional trigger information, more precise timing) is generated
if necessary by routing the relevant signals from the sources to spare FECs and
integrating the resulting data into the data stream exactly as if it were coming
from the PMTs. Interpretation of the data is left to the offline
analysis.

The eCPU reads the current configuration from a control block in the
MDPM that has been loaded by SHARC during initialization. It
loops over the desired hardware, incrementing error registers in the
MDPM if read or write errors are detected, and attaching information
identifying the bad crate or card. The eCPU periodically checks the
configuration block for changes, which if detected will alter the
readout loop. In this manner, individual cards or even entire SNO
crates can be removed from the readout loop. A MDPM-resident control
structure containing ``start'' and ``stop'' semaphores is also polled
periodically, and enables global control of the readout.

Removing hardware from the readout loop or pausing the readout
entirely does not prevent data from being stored in the FEC
memory. Each FEC holds up to $\sim 350,000$ PMT hits, several hours of
data under normal detector operation. Individual channels (bad PMTs)
are taken offline through software by writing to a FEC register that
prevents the channel data from going into the FEC memory.

The eCPU transfers the FEC and MTC memory contents to two circular
buffers in the SDPM, 7~and 1~MByte in size,
respectively. The eCPU updates a buffer write pointer, while the Sun
updates a buffer read pointer, enabling the eCPU to track the Sun's
readout progress. The eCPU will pause the readout of the hardware if
either buffer fills up and will wait until the Sun catches up.

Whenever it is changed, a copy of the configuration control block is
written to a block in the SDPM by the eCPU,  where it is read by
the Sun and incorporated into the event data stream.  

The Sun, once it has read out the SDPM, sorts through the MTC and FEC
data and aggregates PMT hits plus MTC trigger information
into events.  Events are stored in a buffer for some period of time to
ensure that any late PMT hits can still be associated with the proper
event. Once the latency period expires, the events are passed to a
FORTRAN process, which translates the data into ZEBRA
format~\cite{daq:zebra} for use by the offline analysis and writes
the resulting files to disk.

In order to monitor the state of the eCPU code, the MDPM control block
contains, in addition to the start-stop semaphores, a pair of
``heartbeats,'' counters that are incremented in turn when the eCPU is
looping over the readout, or in a tight loop waiting for the start
semaphore to be set. The user interface for SHARC displays
these heartbeats.

Data from the Sun, either in raw, pre-event-built form or as completed
events, are transmitted over the network to a Sun resident process,
called the Dispatcher, which rebroadcasts the data to any clients
requesting it. As this rebroadcast is also over the network, clients
can be located anywhere.  The Dispatcher also receives data regarding
run conditions from the control Mac.  The data are labelled, and
different clients can request data with different labels.  Clients
receive only as much data as they can process, and any data unsent by
the Dispatcher is overwritten after some time.  For high-rate data,
{\it e.g.,} certain types of calibrations, the monitor system behaves
as a sampling system.  The system is extremely modular, and clients on
various platforms can connect and disconnect from the Dispatcher at
will.  Because of this separation between the Dispatcher and the
monitors, the Dispatcher itself is very stable.

Tools available for both UNIX and Mac
platforms  allow one to examine the raw PMT data, MTC data, or
full event data, to check for data integrity, and to examine data
either on a crate-by-crate basis or for the full complement of
channels. There are several tools for event display in the SNO
detector geometry. In addition, there are Web-based monitoring tools
that update HTML files on a designated server for displaying more
static information of interest to the offsite collaborator.

\section{Neutral Current Detector Array}
\label{sec:NCDs}

The detection of a free neutron is the signal that a neutral-current (NC) event
has occurred, while the detection of a neutron (or two neutrons) in
coincidence with a positron indicates a $\overline{\nu}_{e}$
interaction.  Heavy water is an excellent moderator, and a number of
possible strategies for detecting  neutrons can be devised.  One
such strategy, discussed in Sections~\ref{sec:introduction}
and~\ref{sec:water-systems}, involves the addition of a chloride salt
to the D$_2$O.  When a neutron captures on 75\% abundant $^{35}$Cl, it
emits 8.6~MeV in $\gamma$ rays, which mainly Compton scatter.  The resulting
\v{C}erenkov radiation emitted by the electron can be detected by the PMT array in the same
way CC events are detected, and the \v{C}erenkov-light patterns
produced by $\gamma$ rays distinguished from those of electrons on the
basis of their greater isotropy.

Alternatively, neutrons can be detected with high efficiency in
$^3$He-filled proportional counters.  An array of such counters, with
a total length of 770~m, and constructed of highly purified
materials, has been designed for SNO~\cite{NCD:NCDProp92}. In this
way, the NC and CC events can be recorded separately and distinguished
on an event-by-event basis.  Event-by-event sorting makes optimum use
of available statistics, and simplifies interpretation of sparse
signals, such as from a distant supernova.

Detection of thermal neutrons by the reaction $$^3{\rm He} + {\rm n}
\rightarrow {\rm p} + {^3{\rm H}} + 764 \ {\rm keV}$$ takes advantage
of an enormous cross section, 5330 b, and is a well developed art that
has included applications in neutrino
physics~\cite{NCD:Pa79,NCD:Gu89,NCD:Vi94}. Nonetheless, the design of
an array of $^3$He proportional counters for SNO poses unusual
problems.

The neutron production rate in SNO is expected to fall in the range 6
to 42 neutrons per day based on the measurements by the
SuperKamiokande Collaboration~\cite{NCD:SuperK} of the rate of $\nu$-e
scattering, but not the flavor composition.  The rate predicted by
recent solar-model calculations~\cite{introduction:SSM} is 13 per day
when the cross-section calculated by Kubodera and Nozawa~\cite{NCD:KN}
is used.  The neutron signal must not be overwhelmed by backgrounds
from naturally occurring radioactivity in construction materials, nor
can substances be allowed to leach significantly into the heavy water
and interfere with water purification. Helium permeates readily
through many otherwise acceptable materials.  The counters are to
survive under water at absolute pressures up to 3.1~atm for ten years
or more, placing stringent demands on detector longevity and
stability. A counter array with high neutron efficiency is needed, but
not at the cost of significantly obscuring the \v{C}erenkov light from
CC events.  These conditions severely constrain detector design, the
materials employed in constructing the detector array, and the methods
of handling and deployment.

Backgrounds of several kinds can be present.  Most serious is
photodisintegration of deuterium by gammas above 2.22~MeV, as the
neutrons produced are indistinguishable from the $\nu$-induced signal.

At the bottom of the natural Th and U decay chains are two
sufficiently energetic gammas.  Many cosmogenic activities also have
high energy gammas, but only one, 78-d $^{56}$Co, has a long enough
half-life to be relevant.  Energetic photons can also be produced by
capture of neutrons and by ($\alpha$,p$\gamma$) and
($\alpha$,n$\gamma$) reactions.

Alpha particles emerging from the wall of the proportional counter can
leave the same amount of ionization in the gas as a $^3$He(n,p)T
event, as can electrons from $\beta$ decay and Compton scattering.
Even very low-energy events, such as the decay of tritium, can
compromise the signal via random summing.

The principal defense against these backgrounds is to minimize
radioactivity in construction materials and to assure a high degree of
cleanliness during assembly.  To supplement those measures, techniques for
pulse-shape discrimination and position encoding are used, both of which require
digitizing the pulse ion-current profiles event by event.
 Spurious pulses from high-voltage breakdown and
electromagnetic interference must be controlled. Backgrounds of all
types must not only be minimized, but in addition there must be robust
techniques to measure them {\it in situ}.

Because of the efficiency of heavy water as a neutron moderator and
the high cross-section for neutron capture on $^3$He a rather sparse
array of neutral-current detectors (NCD) is sufficient. An array of
5.08~cm diameter proportional counters with a total length of 770~m
arranged on a square lattice with 1~m spacing gives a neutron capture
efficiency of approximately 37\% (see Table~\ref{table:NCD_capture})
with tolerable interference to \v{C}erenkov light produced via the
charged current interaction.  Approximately 15\% of uniformly
generated optical photons are absorbed.
\begin{table}[tb]
   \caption{Capture percentages of neutrons by isotope, for pure $\DO$ and for
            pure $\DO$ with NCDs installed.}
   \label{table:NCD_capture}
   \medskip
   \begin{tabular}{|l|l|l|} \hline
                                           &  Pure $\DO$    &   NCDs \\ \hline
      $^3$He                               &                &  36.6  \\
      Ni                                   &                &   1.6  \\
      $\DO$                                &  66.9          &  36.9  \\
      \hspace*{1cm} H in $\DO$             &  16.5          &   9.1  \\
      \hspace*{1cm} D in $\DO$             &  32.0          &  17.7  \\
      \hspace*{1cm} O(n,$\gamma$) in $\DO$ &   5.8          &   3.2  \\
      \hspace*{1cm} O(n,$\alpha$) in $\DO$ &  12.6          &   6.9  \\
      Acrylic                              &  28.4          &  20.9  \\
      $\HO$ exterior to AV                 &   4.7          &   4.0  \\ \hline
   \end{tabular}
\end{table}

At modest pressures of 1-3~atm of $^3$He, the counters are essentially
``black'' to thermal neutrons. For 99.85\% enriched heavy water, the
mean distance from the point of generation to the point of capture for
thermal neutrons is calculated to be 110~cm and the mean time to capture 16~ms
with the NCD array, compared with 48~cm and 4~ms, respectively, for 0.2\%
MgCl$_2$.

The detectors are filled with a gas mixture of 85\% $^3$He and 15\%
CF$_4$ at a total pressure of 2.5~atm to provide a good compromise
between gas gain and stopping power (to mitigate a ``wall effect''
wherein either the proton or the triton strikes the wall before the
end of its range). A lower pressure, and therefore operating voltage,
would simplify microdischarge management and increase drift speeds,
but would require thick-walled counters to resist collapse.  At the
bottom of the SNO vessel, the absolute pressure is that of a column of
D$_2$O of height 16.8~m plus 1.3~atm of air pressure, for a total of
3.1~atm.

The copper anode wire of diameter 50~$\mu$m is low in radioactivity
and ohmic losses. At 1650~V the gas gain is approximately 100.  For
higher gas gains positive-ion space charge at the wire causes the
avalanche multiplication to depend increasingly on track orientation.

Trace amounts of electronegative contaminants such as oxygen and water
degrade severely the performance of the gas in a sealed
counter. Consequently, counter surfaces are electropolished,
acid-etched, baked under vacuum, and purged with boiloff N$_2$ prior
to fill.  The chemical treatment also removes surface debris stripped
off the mandrel during the fabrication process.

The $^3$He gas from the Department of Energy facility in Savannah
River contains a small amount of tritium, about 0.5~mCi/l, which is
reduced to 5~nCi/l or less by passage through a charcoal-loaded cold
trap and by recirculation through a SAES, Inc.,  St101 getter.  At that level,
random coincidences and pileup of tritium decay pulses are no longer a
concern in the energy regime of interest.  Some detectors, about 5\%
of the total by length, are filled with $^4$He:CF$_4$ to provide a
check on backgrounds.

\begin{figure}[ht]
   \setlength{\epsfxsize}{0.5\textheight}
   \centerline{\epsffile{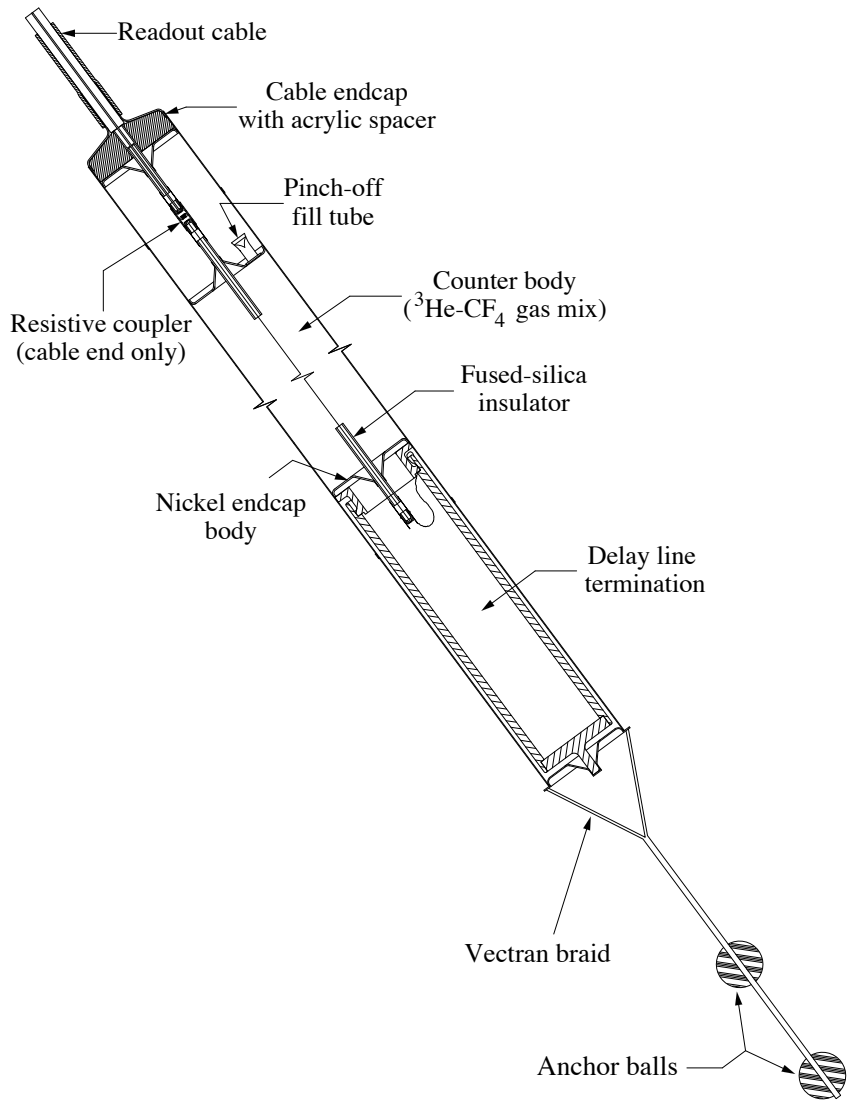}}
   \caption{SNO neutral-current detector.}
   \label{fig:NCD_det}
\end{figure}

The NCDs are fabricated in three different unit lengths in order to
fill the sphere efficiently.  ( An upper limit is set by the
dimensions of the cage in the INCO Number 9 shaft, which permits an
object 3.7~m long to be brought down.)  The active length is 13~cm
less than the mechanical length.  Fig.~\ref{fig:NCD_det} shows the
main features.
The important mechanical parameters of the detectors are summarized in
Table~\ref{table:NCD_Mechanical}.
\begin{table}[tb]
      \caption{Important physical parameters of the NCD detectors.}
      \label{table:NCD_Mechanical}
      \medskip
                                                \begin{tabular}{|l|l|l|l|}
                                        \hline
         Item           & Value             & Variation            & Units
\\ \hline
         Wall thickness & 0.036             & $^{+0.012}_{-0.000}$ & cm \\
         Length         & 200, 250, and 300 &                      & cm \\
         Diameter       & 5.08              &                      & cm \\
         Wire diameter  & 50                &                      & $\mu$m
\\
         Gas pressure   & 2.50              & $\pm$0.03            & atm \\
         Weight         & 510               &                      & gm/m
\\ \hline
      \end{tabular}
\end{table}

The bodies of the proportional counters are made of ultrapure nickel
tubes fabricated at Mirotech (Toronto) Inc., by thermolysis of
Ni(CO)$_4$ vapor at the surface of an anodized aluminum mandrel heated
to 215~C. A limited number of elements (Pb, Ra, Th, and U not among
them) react with CO to form carbonyls, and the metals formed by
chemical vapor deposition (CVD) from this precursor can be expected to
be free of the most troublesome radioisotopes.  The nickel deposit has
properties very similar to conventional metallurgical products.
Radiochemical neutron activation analysis has shown Th levels in the bulk
material of order 10$^{-12}$ (1 ppt) by weight or less.

Nickel is a relatively inert metal.  The exposed macroscopic surface
area in SNO is 125~m$^2$.  In 17.3~M$\Omega$-cm water the measured
leach rate is $\leq$22~$\mu{\rm g \ }{\rm m}^{-2}{\rm d}^{-1}$, which is
not expected to compromise the operation of the water purification
plant.

Endcaps are also made by CVD, in this case on stainless steel
mandrels, and are welded into the tubes with a Lumonics 50-W Nd-YAG
laser welder.  Insulators are Heraeus-Amersil Suprasil T-21 synthetic
fused-silica tubes.  The insulators are internally coated with a layer
of pyrolytic graphite at anode potential to eliminate electric fields
inside them.  They extend 2.5~cm into the gas volume to act as field
tubes and prevent multiplication of electrons from regions where the
electric field is distorted.  A silica-nickel seal is highly
mismatched, so the design places the silica under compressive stress
at working temperatures to take advantage of the high compressive
strength of that material.  Techniques were developed in collaboration
with IJ Research (Santa Ana), Inc. to metallize the seals and solder
them with a 96.5:3.5 eutectic Sn:Ag alloy.  Counters are filled
through copper tubes, which are then pinched off.  All assembly is
carried out in a Class 1000 cleanroom.

The individual units are  welded
together at the time of deployment in SNO in 96 ``strings'' of length ranging
from 4~m to 11~m. A total mechanical length of 770~m is to be deployed in the
heavy water. The strings are installed by a remotely operated vehicle (ROV) once
the vessel is filled with heavy water.  Each string is anchored to an
attachment point affixed to the bottom of the acrylic vessel and
floats upward, restrained by a flexible braided anchor cord made of
Vectran fiber.

Spaces of 35 to 50~cm are left between the string ends and the
vessel. The bottom of each string is terminated with a 30~ns
open-ended 415$\Omega$ delay line to facilitate position readout by
pulse-reflection timing.  A single 91$\Omega$ coaxial cable, made of
copper and polyethylene by South Bay Cable (Idyllwyld, CA), Inc.,
carries signals from the top of each string up the neck of the vessel
to preamplifiers.  The cable specific gravity is 0.955.

In Table~\ref{table:NCD_photo} are listed some of the decay properties of
relevant isotopes. Since detection of \v{C}erenkov light is the
principal means of quantifying the photodisintegration background {\em
in situ}, the production of light by other activities is also
important.

\begin{table}
\caption{Photodisintegration and high-energy beta emitters.}
\label{table:NCD_photo}
\medskip
\begin{tabular}{|l|l|l|l|l|}
\hline
Isotope & E$_{\beta{\rm (max)}}$, MeV &  E$_{\gamma}$, MeV & $\gamma$ branch,
\% & ($\gamma$,n) Probability \\
\hline
$^{212}$Bi($\ra ^{208}$Tl) & 1.8 (18\%)  &  2.615 & 36\%  & 1/470 (per $\gamma$) \\
$^{214}$Bi                 & 3.26 (18\%) &  2.445 & 1.5\% & 1/750 (per $\gamma$) \\
$^{56}$Co                  & 1.8         &  var.  & -     & 1/1125 (per decay)  \\
$^{228}$Ac                 & 2.06 (11\%) &  var.  & 89\%  & 0  \\
$^{234m}$Pa                & 2.28 (100\%)& -      & -     & 0  \\
$^{40}$K & 1.3 (89\%)      & 1.46 (EC)   & 11\%   & 0  \\
\hline
\end{tabular}
\end{table}

By means of a suite of radioassay techniques (neutron activation
analysis, radiochemical methods, direct gamma counting, alpha
counting) applied to samples and to the complete inventory of smaller
components, quantitative predictions can be made concerning the levels
of photodisintegration and other backgrounds that will accompany the
NCD array in SNO.  A neutron production rate of 100 per year in the
SNO vessel results from 1.1~$\mu$g of Th in equilibrium, or
12.8~$\mu$g of U in equilibrium. Slightly more than half of all
neutron-capture events can be unambiguously identified via track
length {\it vs} energy as being distinct from alpha particles (for
this reason the array efficiency is taken to be 25\% rather than the
nominal 45\% capture efficiency).  Electron and Compton backgrounds,
and microdischarge events, have topologies separated still further
from neutron events.  As a result, photodisintegration is the only
background requiring a separate determination and subtraction, which
can be accomplished by measurement of the \v{C}erenkov light emitted
by radioactivity.  Assays still in progress preliminarily indicate
that the NCD array will contribute fewer than 300~neutrons per year.

\section{Control--Monitor--Alarm (CMA) System}
\label{sec:CMA}

The Control--Monitor--Alarm (CMA) system is the data-acquisition and
control system for the underground laboratory environment. The system
monitors approximately 250 analog and 220 digital readings. Major
functions of the CMA system are:
\begin{itemize}
   \item Hazardous gas detection;
   \item Environment monitoring (temperature, humidity, air pressure);
   \item Cooling and air-circulation control;
   \item Load and position sensing on critical structural elements;
   \item Electrical system monitoring;
   \item Control of the magnetic compensation coils for reduction of the effect of the earth's
field;
   \item Monitoring of supply voltages and currents for the electronics;
   \item Logging of various monitored parameters.
\end{itemize}
The system uses Paragon TNT industrial automation software from
Nemasoft~\cite{cma:paragon}. The hardware was furnished as a complete system
by Sciemetric Instruments~\cite{cma:sciemetric}.  Analog readings are
converted by a single multiplexed 20~kHz 12-bit ADC. Voltages up to
$\pm 10$~VDC can be converted in seven software selected ranges. Industrial
4--20~mA signals are converted to voltages with precision resistors.  Digital
inputs may be 24~VDC or 120~VAC. There are 32 solid-state relay channels for
digital output. The input/output scan rate is up to 2~Hz. Systems with
independent microprocessors (such as the cooling unit, and the
uninterruptable power supply) have serial links for ASCII output to the CMA.
  
The CMA software is presently running on a dual 200~MHz Pentium-Pro
system with 256~MBytes of RAM running Windows NT 4.0. Remote systems
running Paragon TNT software can connect via TCP/IP with the
underground system. A remote monitoring system is kept running in the
SNO surface building. Critical alarms are annunciated on the surface
as well as underground.

Data are recorded continuously in a Microsoft Access database on the
main CMA system, and are exported periodically to the custom SNO
database. The data logging program enters a new record only when a
significant change in a parameter has occurred.

\section{Calibration}
\label{sec:calibration}

Accurate measurement of the physics processes in the SNO detector
requires a chain of calibrations and calculations to connect the raw
measured quantities of PMT charge and time to a full description of
the interaction in terms of energy, direction, and particle type. This set of
calibrations includes built-in electronic calibrations, several dedicated
optical calibrators, various types of tagged and untagged radioactive sources
and detailed radioassays of the detector materials. Taken together with a
comprehensive Monte Carlo simulation of the detector,
this set of calibrations creates an overdetermined model of the detector that
can be used to interpolate energy scales and particle identification
across the full response range of the detector. The major calibration tools are
listed in Table~\ref{table:calibration_categories} and  briefly
described in the following sections.

\begin{table}[htbp]
\caption[]{Calibration devices for the SNO detector and their principal uses.}
\label{table:calibration_categories}
\begin{tabular}{|p{4cm}p{8cm}|}
\hline
Device & Calibrations \\
\hline \hline
Electronic Pulsers & Time slope (ns per count);
		     Time pedestal;
                     Charge slope (pC per count);
		     Charge pedestal \\

\hline
Laser Ball; Sonoluminescent Source; LED Sources
                 & Common time reference ($t_0$); 
                   D$_2$O absorption and scattering ($\lambda$);
                   Acrylic absorption and scattering ($\lambda$,r,$\theta$);
                   H$_2$O absorption and scattering ($\lambda$);
                   Reflection and scattering from structures;
                   Single photoelectron (pe) response ;
                   Relative single pe efficiency;
                   Multiple photoelectron response;
                   Walk (time vs. pulse amplitude) \\
\hline
Radioactive Sources: $^{16}$N, $^{252}$Cf, $^8$Li, Th, U
                 & Gamma energy response;
                   Electron energy response;
                   Neutron-capture efficiency;
                   Angular response \\
\hline
$^3$H(p,$\gamma$)$^3$He Accelerator Source 
                 & Gamma energy response \\
\hline
Tagged Sources: $^{252}$Cf, $^{24}$Na, $^{228}$Th
                 & Gamma energy response;
                   Neutron-capture efficiency \\
\hline
Radioassay       & Backgrounds from construction materials;
		   Backgrounds from  water contaminants  \\
\hline
\end{tabular}
\end{table}

The conversion of the digitized values for PMT charge and time to
photoelectrons (pe) and nanoseconds (ns) is a two-step process,
involving electronic pulses and light flashes from known positions
within the detector. The electronic pulser calibration yields the transfer
functions and offsets (pedestals).  The 
establishment of a common time reference for all channels ($t_0$) is performed
with optical sources.

The charge offsets (pedestals) are the values digitized when there is
no charge input from the PMT.  The time offset is found using a full
time ramp.  During calibration, each cell (there are 16 analog memory
cells per channel for each of the three charge and one time
measurement) is given 10 event trigger pulses without any PMT input.
Because the pedestal calibration is done during normal running there
are occasional coincidences with noise or detected light that must be
removed during the analysis stage.  In the offline analysis, two
passes are made through the data.  The first pass determines the value
of the digitization that occurs most frequently, and then events that
are more than $\pm 10$ units away from this peak digitized value are
discarded for a second pass that determines the mean (pedestal) and
RMS for each cell.  Should more than a certain number of events be
rejected with this cut, the cell is flagged and not used again until
recalibration or repair.  The RMS is typically less than one ADC count
and cells with an RMS greater than three counts are flagged.  The
pedestals are run constantly in background while solar-neutrino data
are being taken.

Individual voltage pulses of varying width and amplitude (and, thus,
charge) are fed to the PMT inputs to measure the high gain,
long integration (QHL), high gain, short integration (QHS), and low gain,
selectable integration (QLx) signals.  Each cell is pulsed ten times for a
given injected charge (\qinj).  The charge injection is incremented in steps up
to a maximum value.  A low-order polynomial fit of digitized value
vs. \qinj\ is performed for each cell offline.

The time calibration sends one pulse to the PMT input and a delayed
pulse to the event trigger; the time difference between the pulses is
varied by a known amount in steps as small as 100~ps. The slope
calibration procedure is similar to that of the charge
calibration. The relation between the time digitized value and
nanoseconds between two generated pulses is monotonic over most of the range. 

The risetime of the PMT pulses and leading-edge discrimination lead to a timing
dependence on the PMT anode charge.  Calibration of this effect is
made by running the laserball at various intensities from a fraction
of a pe per PMT to hundreds of pe per PMT, and then creating charge
versus time (QvT) histograms for each PMT.

\begin{figure}[t]
   \setlength{\epsfxsize}{0.75\textheight}
   \centerline{\epsffile{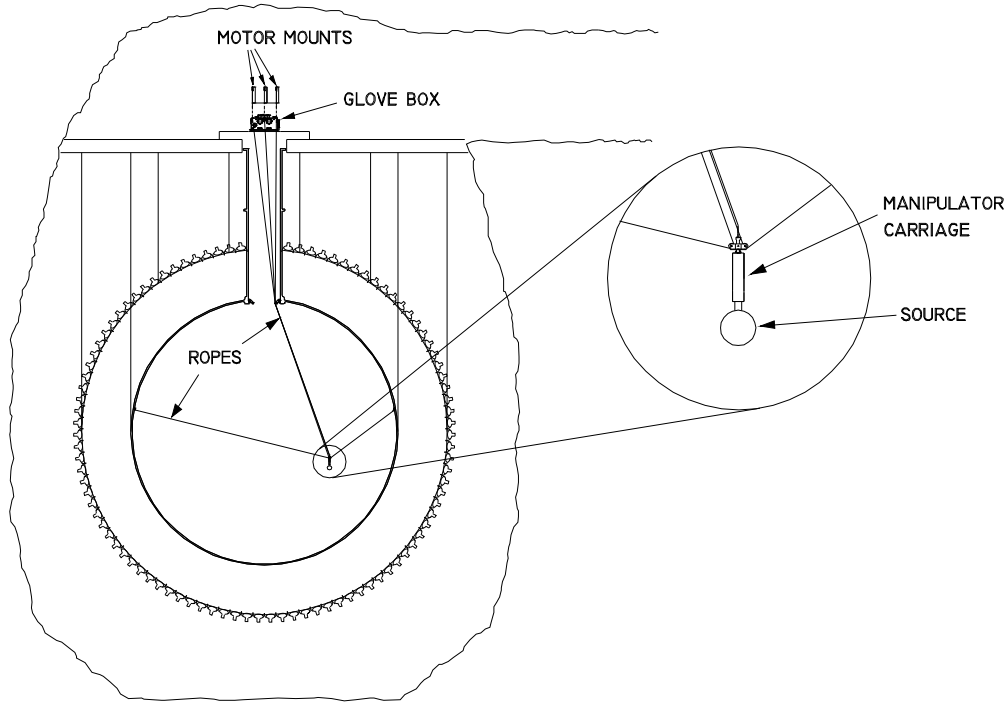}}
   \caption{SNO calibration source manipulator system.}
  \label{fig:calibration_manip}
\end{figure}

Because many of the optical- and particle-calibration techniques are
position dependent, a general-purpose manipulator has been
developed to provide source deployment at accurately known off-axis
positions in two orthogonal planes in the \DO.  (In addition, six
calibration access tubes run from the deck to positions in the light water.) 
The source manipulator is a carriage located in a plane by two Vectran ropes
(see fig.~\ref{fig:calibration_manip}) that run down the vessel neck
through pulleys on the carriage to fixed locations on the acrylic
vessel. A third, center rope connects directly to the source
carriage. The source can be introduced at the neck and visit 70\% of
the plane with a position accuracy of better than 5~cm.

At the deck interface the carriage is pulled up into a glove box that
is part of the cover-gas volume.  There a person can work on a source
while keeping the detector isolated from light and airborne
radioactivity.  The source can then be pulled through a gate valve
into an interlock box where it can be removed or serviced.

Some of the calibration sources require power cable, signal cable
and/or gas capillary connections to equipment on deck. To provide
uniform and reliable manipulator operation, these auxiliaries are
cabled into a standard sized umbilical cable.  The umbilical cable runs through
the center of the source carriage to a compression fitting at the top
of the source. The cable connections are then internal to the source.
The umbilical cables are stored and paid out from a pulley array mechanism
held under tension, thereby avoiding twisting of the umbilical cable.

Optical sources are used in both the optical calibration and the
electronics calibration. They include a wavelength-selectable
dye laser, a sonoluminescent source, and light-emitting diodes
(LEDs). Both the laser and sonoluminescent source have emission times
 well suited to making precision timing calibrations. The LED
sources are slightly broader in time, but multiple sources in
well-surveyed positions offer a useful and easily operated check on
the detector geometry and stability.

The nitrogen or dye laser with a fiber-optic cable to a diffuser ball
(``laserball'') is the primary light source~\cite{calibration:laser}.
The laser (Laser Photonics Inc. LN203C) provides pulsed radiation at
$337.1$~nm with a $600$~ps pulse width and peak power of $150$~kW.
The laser pulses up to $45$~Hz and can be used directly or as a pump
for several dye lasers that provide wavelengths in the range of
360--700~nm. Neutral-density filters mounted into two selector
wheels allow remote intensity adjustment. The laser light is delivered
to a diffuser ball via a fiber-optic bundle within an
umbilical cable.

The dye laser frequency and attenuation filters can be adjusted
remotely by computer control.  The output from the dye lasers is
focused into quartz fiber with length 1~m and diameter 1~mm (Fiberguide
Ind. SFS-1000N) to enhance beam structure stability.

The laser pulses are transmitted to the detector by a bundle of 22
high-OH $100$~$\mu$m silica step-index fibers (Fiberguide
Ind. SFS-100/110/240/310N). These fibers offered the best combination
of UV transmission, low dispersion and cost.  The fiber-optic
umbilical cable terminates with a 10~cm diameter diffuser ball that
serves to distribute light quasi-isotropically through the detector.
A plug at the neck of the ball contains a fiber-optic positioning
guide tube and also houses a miniature monitoring PMT (Hamamatsu
R5600).  The fiber-optic guide tube allows the umbilical cable to be
detached from the laserball and reconnected with no effect on the
light output pattern. The light distribution pattern from the
laserball is mapped out with monitor PMTs by rotating the ball in a
dark room.  The diffuser ball light pulse width is about $2\,$nsec
(FWHM).

A sonoluminescent source provides bright, isotropic light of extremely
short duration. Single bubble sonoluminescence (SBSL) can occur when
an air bubble is introduced into water at the node of an intense,
spherically symmetric sound field. The light\cite{cal:Hilgenfeldt} has a very
short ($<100$~ps) pulse width and a frequency distribution corresponding to a
blackbody at a temperature of over 20,000~K~\cite{calibration:sono_report},
very similar to the
\v{C}erenkov spectrum. The relatively high repetition rate 
($\sim$25~kHz) offers a good test of the system's burst capability.
The short pulse width and isotropic output make the SBSL source useful
for PMT timing calibrations and some optical analysis.  In addition,
the SNO detector provided an opportunity to investigate a variety
of fundamental properties of SBSL~\cite{calibration:sono_dougm_thesis}.

The SNO SBSL source consists of a 250~ml spherical flask with two
cylindrical piezoelectric transducers.  The flask is housed within a
sphere made of two 15~cm radius hemispherical acrylic sections to
provide an air buffer. It is filled with pure distilled water to the
neck level, and driven at the natural monopole resonance (28.5~kHz)
with an inductively matched circuit.  A small microphone transducer
aids in scanning the frequency to find the natural resonance.

A bubble is generated remotely with a nichrome heater wire within the
water.  The bubble stability is monitored by observing the spikes on
the microphone voltage trace. As sonoluminescence output begins, the
light output is monitored with a small PMT (Hamamatsu R5600) within
the acrylic source housing. The light output contains of order a
million blue photons which are isotropically distributed in space when
averaged over time scales of seconds or longer.  The heater wire also
provides some control over the light intensity, which decreases with
increasing water temperature.

Light emitting diodes provide a simple and robust optical source for
verification of water transparency, acrylic vessel and source
positions, and PMT response. During the final construction stages of
the PSUP, new LED products became available from Nichia Corporation.
These indium-gallium nitride single quantum well (NSPB500) and double
hetero-structure (NLPB500) devices offer high intensity blue radiation
(420--480~nm), and the latter device can even emit in the UV (380~nm) when
strongly overdriven in pulse mode~\cite{calibration:blueleds}.  A
pulse of width approximately 7~ns is observed. Six of these devices
have been packaged with a pulser circuit and mounted at the locations,
in cm, (695, 248, 399), (-162, 681, -463), (-699, 49, -462), (-8, -764,
-346), (698, 54, -463),  and (37, 1, -838) on the PSUP, where the
center of the PSUP is (0,0,0), the z axis is up, and the x and y axes are
defined with respect to construction coordinates.

The optical calibration process results in the determination of
attenuation and scattering coefficients for the heavy water, light
water, and acrylic~\cite{calibration:optical_cal}.   By using the various
measured optical coefficients, it is possible to simulate via a Monte Carlo
calculation the optical transport of \v{C}erenkov light produced by a neutrino
or background particle. These simulations can then be used to interpolate (in
space and amplitude) between the various energy calibrations discussed in
the next session.

Most of the optical analysis is performed using data from the
quasi-isotropic laserball at near single photoelectron (spe) light
intensity, and utilizes the detected light intensities as a function
of the calibrated time between the fast laser trigger and the
digitized time value of the PMT discriminators.  Light reflected from
various parts of the detector produces substructure in the time
distribution.  This substructure can be used to characterize the
optical properties of various parts of the detector, such as the PMT
reflectors and the acrylic vessel, and to determine the Rayleigh, Mie
and large particulate contributions to the scattering of light in the
water
volumes~\cite{calibration:scattering,calibration:ecoli_scattering}.

By moving the laserball to various off-center positions, absorption
coefficients can be measured.  Calibration source access ports into
the \HO\, volume within the PSUP will aid in determining the light
attenuation in the acrylic panels.  The relative PMT anode
efficiencies and the overall optical ``gain'' of the detector are to
be monitored at the spe level over time, and can be separated from
changes in other optical constants by taking data at several source
positions and at $400$~nm where the acrylic transmission is high and
uniform across acrylic sheets.

There are three gas transport sources: a $^{16}$N gamma ray source, a
$^{17}$N neutron source and a $^8$Li beta
source~\cite{calibration:gas_cal}, of which the $^{16}$N is fully developed.
The gas transport sources consist of a D-T generator neutron source, a
neutron target chamber, gas capillaries, and decay chambers that can
be deployed within the detector. The D-T generator, producing 14~MeV
neutrons, is enclosed in concrete 30~m from the detector. Various
target chambers can be moved into the neutron flux by stepper-motor
control. Gases from the target chamber are routed via capillaries to
the deck area and down into the source umbilical cables. The decay chambers
can attach to the manipulator carriage and connect to the umbilical cable
gas lines.

The $^{16}$N is produced via $(n,p)$ on the oxygen in CO$_2$ gas and
decays by a beta delayed 6.13-MeV (66\%) or 7.12-MeV (4.8\%) gamma ray
with 7.13-s half-life.  The decay chamber is 41.9~cm long and 5.7~cm
radius, and is made from 5-mm thick stainless steel so that the
electrons (which have a 10-MeV endpoint energy) are absorbed.  A beta
tag is provided by a 3-mm thick scintillator sleeve on the inside of
the chamber with a 5-cm PMT. The source has been tested in the SNO
detector and produced over 300 $^{16}$N decays per second.

To reach gamma energies above 6~MeV a miniature accelerator source drives the
$^3$H(p,$\gamma$)$^4$He reaction to produce 19.8~MeV
gammas~\cite{calibration:pt_accel}.  The source consists of a
cold-cathode Penning ion source, electron optics for beam extraction
and acceleration, and a fixed target. It is housed in a 25-cm diameter, 60-cm
long stainless steel capsule and operates with a beam voltage of
20--30~kV to produce approximately one photon per second.

Tagged sources have been developed that permit specific kinds of
events in SNO to be identified by time coincidence with a trigger,
allowing one to acquire calibration data in an essentially
background-free manner.  Such sources are valuable for testing PMT
response, particularly at low energies, for determining backgrounds,
for checking neutron transport simulation code, and for calibrating event
reconstruction.

Tagged neutron sources are fabricated using $^{252}$Cf at several
different activity levels. The 2.64-y $^{252}$Cf decays with a
spontaneous fission branch at 3.09\%, which provides neutrons and a
trigger from fragment betas and gammas.

Another tagged source is constructed from a NaI crystal coupled to a
PMT.  The crystal is activated to produce 15-h $^{24}$Na.  The trigger
can be gated on the energy deposited in the NaI in order to tag the
1.37~MeV or the 2.75~MeV gammas in SNO to study the response to
low-energy gammas, and to measure photodisintegration neutrons.  A
third source~\cite{calibration:pc_source} under development is a
polycarbonate-housed proportional counter with a radioactive source
({\it e.g.}, $^{228}$Th) deposited on the anode wire.  The source is
designed so that emitted betas will deposit little energy in the
counter and will mimic the spectrum produced by natural radioactivity
present in SNO.

After correcting for relative response due to position and direction,
an event is assigned an energy according to a calibrated NHIT scale.
The $^{16}$N source  sets the absolute scale and provides a
measurement of the energy resolution for a 6.13-MeV gamma ray.  The
energy response to electrons will be inferred from this measurement
since the difference in response between gamma rays and electrons is
well modeled with the EGS4 code and will not contribute significantly
to the uncertainty in the electron energy scale.  For each event there is a
significant probability for recording unrelated noise hits. The rate
of background hits is directly measured by regularly sampling the
detector using the pulsed--GT trigger.  The high-energy end of
the scale is calibrated with the accelerator source.

The detector angular resolution is primarily determined from
theoretical treatment of electron scattering and production of
\v{C}erenkov light, together with the geometry of the PMT array.  The
treatment can be verified through the Monte Carlo against data from
the $^{16}$N gamma source.  If the fitted event position is taken to
be the point of momentum transfer from the gamma ray to the electron,
then the initial direction of electron motion for single Compton
interactions is the vector to this position from the $^{16}$N source
position.  A cut on large radial distances reduces the source angle
uncertainty, while a cut on the larger NHIT events will select mostly
those events of large momentum transfers to the electron in the first
scatter.  Deconvolution of the Compton scattering angle can be
accomplished with the Monte Carlo simulation.

\section{Control of Radioactive Contaminants}
\label{sec:cleanliness}

The minimization of background events requires a high intrinsic purity
of materials and the reduction of radioactivity on the surfaces of the
detector components. The purity of materials was addressed at the time
they were selected for fabrication by measuring the activity in
samples.  Low background $\gamma-$ray counting and neutron activation
analysis have been used extensively for analysis of different
materials in the SNO detector.  Since the detector is located in an
active mine where the rock and ore dust contains typically 60~mg/g Fe,
about 1.1~$\mu$g/g U, and about 6.4~$\mu$g/g Th, while the inner
components of the detector require at or below $10^{-12}$~g/g U and
Th, dust levels had to be kept very low.  A program to keep
contamination as low as reasonably achievable during the construction
and operation of the detector was planned from the inception of the
experiment. This section describes the methods for controlling and
reducing surface contamination.

Radon ($^{222}$Rn) is of special concern in the SNO detector, since it
can diffuse from components, dust or the cavity walls into the water
and move into critical regions of the detector. Radon emanation from components
depends on diffusion properties as well as the U content, and hence it varies
with the density, hardness and surface properties of the materials.  Extensive
 measurements of radon emanation were carried out in both vacuum and
water test chambers for all critical materials for the SNO detector.

Because of the presence of small amounts of radioactive contaminants--$1.1\pm
0.2$~$\mu$g/g U and $5.5\pm 0.5$~$\mu$g/g Th in the host rock and
$1.2\pm 0.1$~$\mu$g/g U amd $2.4\pm 0.2$~$\mu$g/g Th in the
concrete liner--in the SNO detector cavity, radon ingress from the
cavity walls is a potentially serious background source.  Radon
diffusion studies were carried out on several polyurethane coating
materials, both in the laboratory and in an underground test facility,
to determine the radon attenuation properties of these sprayed-on
coatings.  To meet the design maximum of 12 radon atoms per square
meter per hour diffusing into the cavity from rock, concrete and liner
materials, the coating material (Urylon HH453 and 201-25
polyurethanes) was required to be $8 \pm 1$ mm thick.  The coating was
applied to a smooth concrete base on the cavity walls and floor in 9
separate layers (of alternating white and gray colour).  The coating
thickness was monitored with a portable probe, developed especially
for this work, which measured the amount of backscattered x-rays from
a radioisotope source mounted in the
probe~\cite{cleanliness:thickness_monitor}.  This backscatter signal
was proportional to the coating thickness and a standard calibration
procedure was developed to give coating thicknesses directly.  The
coating is expected to attenuate radon from the walls by a blocking
factor of $2\times10^{-7}$.

Contamination control at SNO has several elements. At the beginning of
assembly of the detector a clean environment (air and surfaces) was
established in the underground laboratory and maintained throughout
construction.  The entire laboratory became, and continues to be, a
clean room.  Fresh air entering the laboratory and the air that is
cooled and circulated within much of the laboratory passes through
HEPA filters.  Detector components were cleaned above ground and
appropriately packaged for shipment underground.  Material and
equipment enter the laboratory through an interlocked area with
high-pressure water cleaning equipment.  Personnel take wet showers
and put on clean garments before entering the laboratory proper.
These transition areas, the car wash and the personnel entry, are
indicated on the general layout of the laboratory (see
Fig.~\ref{fig:LabLayout}).

The surfaces of the laboratory (walls, floors, and construction
equipment) were cleaned repeatedly as necessary during construction.
Personnel pass through air showers at critical locations.  Components
of the detector (such as the acrylic vessel) were given a final
cleaning after assembly.  Components that could not be accessed for
cleaning or that have topologically complicated surfaces, for example,
the PMTs and reflectors, were covered wherever possible with clean
tarpaulins or plastic sheeting (dust covers), which were removed at a
later time.  Dust covers were used extensively in the construction of
the acrylic vessel and the PSUP.  Toward the end of construction, the
infrastructure (platforms, ladder ways, ductwork, elevator, etc.) was
removed in a manner that left the detector cavity clean.  Care was
required as much of this infrastructure was installed before the
cavity was first cleaned.

At the end of construction the cavity deck was sealed to inhibit the
movement of radon, and a separate clean room was established on the
deck above the chimney of the acrylic vessel.  This local clean room
encloses the insertion points and devices that are used in the
calibration and operation of the detector and subsequently in the
installation of the neutral-current detectors.

To control surface contamination one must be able to measure it and to
monitor the factors affecting it. Commercial air-particle counters are
used to sample the number and sizes of fine particles in the air.  The
number of particles larger than 0.5 $\mu$m in diameter in 28 l
of ambient air defines the ``class'' of the air and of the clean room. 
Class values vary dramatically depending on the
work  going on in a particular area.  In general, the quality of the
air within the laboratory has been in the range of class 1000 to
10,000, depending on location and level of activity.  The amount
(mass) of dust that accumulates depends on the number versus size
distribution and on the density of the particulate.  Mass versus size
distributions for mine dust were measured with a cascade impactor, and
number-versus-size distributions with an optical microscope. 

The dust deposition rate and the total accumulated dust are the most
relevant quantities.  Several methods were used to measure the amount
of contamination on a surface.  A quick method is the ``white glove
test,'' in which a white cloth, mounted on the sharp edge of an eraser,
is wiped over a surface for a given distance. Through calibration
measurements and adoption of consistent procedures this test has been made 
 semi-quantitative, and it is sensitive down to  about
0.5~$\mu$g/cm$^2$ of dust. 

X-ray fluorescence spectrometry (XRF) was used for more sensitive and
precise measurements.  In this method, one detects elements in the
range  $Z = 20$ to 45 by the characteristic X-rays emitted following
fluorescence of the sample  with an x-ray tube.  A spectrometer was
built for this purpose and located underground for convenient use.

In one type of XRF measurement, thin, spectroscopically-pure plastic
films were used as ``witness plates'' to collect dust over a known
period of time.  The amount of mine dust deposited was inferred by
measuring the amount of iron, which is present in the local mine dust
at the level of 7\%.  A sensitivity of about 0.1~$\mu$g/cm$^2$ is
possible in a counting time of 20 minutes.

In a second type, dust is removed from a surface by placing a section of thin
adhesive tape (low in Fe) on the surface, pulling it off and placing
it in the spectrometer.  When a single piece of tape is used to
perform five lifts from adjacent areas on a surface, the sensitivity
can be extended down to about 0.015~$\mu$g/cm$^2$.  This method was
used to monitor the surface cleanliness of the acrylic vessel. 

A few direct measurements of U and Th in deposited
dust were made using neutron-activation analysis to verify the
correlation of Fe with U/Th.

During more than three years of construction since clean conditions were
first established, fairly regular patterns and deposition rates
emerged.  The quality of the air was characterized by an average class
value of $2500 \pm 500$. Average deposition rates ranged from
1~$\mu$g/cm$^2$/month of mine dust in busy areas  to one
tenth that amount in inactive areas. (The air particle counts
and mass deposition rates are roughly consistent with the number-size
distributions for airborne particulates measured inside the
laboratory.)  The level of cleanliness on the surface of the acrylic
vessel at the time of water fill was typically better than 0.1~$\mu$g/cm$^2$. 
(Zinc was also regularly observed in the XRF spectra at a level typically 20\%
of the mine dust.  Zinc comes mainly from the use of galvanized scaffolding
in construction.)  Since a mass deposition of 1~$\mu$g/cm$^2$/month integrated
over a thirty-month period would exceed the target contamination limits in any
region of the detector, periodic cleaning of accessible surfaces and the use of
dust covers on inaccessible areas was required to achieve target levels.

\section{Offline Analysis and Simulation Code--SNOMAN}
\label{sec:SNOMAN}

The offline software is required to perform two major functions: the
analysis of SNO data, and a detailed and complete simulation by Monte
Carlo techniques of all significant signals and backgrounds using as
accurate a model of the detector and its response as possible.  Both
of these functions are combined in the SNO Monte Carlo and Analysis
code, or SNOMAN.  SNOMAN consists of a set of largely autonomous
processors written primarily in FORTRAN.  In order to ease distributed
development and maintenance (and to overcome the lack of dynamic
memory allocation in FORTRAN) these processors communicate through a
central data structure managed by the CERNLIB package ZEBRA.  ZEBRA is
also the basis for the management of banks of data, called ``titles.''
Titles-management routines in SNOMAN insure that as successive events
are analysed the constants describing the detector response are the
appropriate ones for that time and detector configuration.  The use of
autonomous processors working on a central data structure allowed
staged development of the processors.  Since at the time of this writing there
are over 25 processors in SNOMAN,  only selected ones will be discussed.

In addition to SNOMAN there is another major offline analysis tool: the SNO
database, or SNODB, which is based on the CERNLIB package HEPDB.  SNODB is a
distributed master-slave database that runs collaboration wide via  PERL
installation and management scripts.  The scripts set up SNODB on the large
variety of platforms in use and manage the updates that keep the various slave
copies of the database in step with the master copy.  SNODB contains
ZEBRA banks that encode all the information about calibrations and
detector status, configuration, and performance which is needed to
interpret SNO data.

The basic units of the data structure for the Monte Carlo are the
vertex and the track.  A vertex represents any event of interest in
the movement of a particle (the origin of the particle, when it
encounters a boundary, any interactions it undergoes, etc.), while
tracks specify how particles move from one vertex to the next.  The
Monte Carlo can then be divided into generation (deciding what seed
vertices to create), propagation (taking the initial seed vertices and
creating successive tracks and vertices based on the transport of that
type of particle in the SNO detector, and creating any additional
particle such as \v{C}erenkov photons which the initial particle may
spawn), detection (either of photons by the PMTs or of neutrons by the
NCDs), and electronics simulation.  In order to achieve accurate,
usable code with the least time and effort well-established and tested
software packages have been used wherever possible. For example the
propagation of electrons, positrons, and gammas is handled by
EGS4~\cite{snoman:EGS4} (modified to produce tracks and vertices), the
propagation of neutrons is handled by routines lifted from
MCNP~\cite{snoman:MCNP}, the muon physics is largely taken from
GEANT~\cite{snoman:GEANT}, and much of the code that describes how the
spectra might be modified by the MSW effect is taken from the work of
Hata~\cite{snoman:Hata}.  Cross-sections doubly differential with
respect to energy and angle for neutrino interactions with deuterium
are those of Kubodera and Nozawa~\cite{NCD:KN}.  A large amount of
work has gone into verifying several aspects of these modified
external codes, which are particularly critical to SNO physics but
which may not have been extensively tested in the past.  In particular
a considerable amount of effort has been expended on understanding the
angular distribution of \v{C}erenkov photons produced by an electron
of a few MeV~\cite{snoman:eleang}, which is important for some of the
analysis techniques discussed below.

The Monte Carlo code for which reliable, pre-tested software like EGS4
was not available was written by the SNO collaboration.  The photon
propagation includes the effects of attenuation, refraction,
reflection, and Rayleigh scattering (Mie scattering is expected to be
negligible for the levels of cleanliness expected in SNO) as a
function of wavelength and polarisation.  The light concentrator/PMT
response has been the focus of substantial software development and
experimental effort.  A detailed stand-alone simulation of a
concentrator and PMT was written~\cite{snoman:lay} and tested
against experiment~\cite{snoman:lyon}, then installed as a SNOMAN
processor.  

The geometry of the detector is a compromise between a totally
data-driven system that would be slow but flexible and a completely
hard-wired structure that would be fast but inflexible.  The geometry
consists of detector elements, each of which is made up of primitives
(such as spheres, cylinders, elliptical toroids, etc.) for which the
actual calculations are made.  The ability to alter the complexity of
the geometry simulation allows the user to switch off features that
are computationally expensive in situations where they do not
significantly affect the results.

The actual data from the SNO detector are written to tape in the form
of packed ZEBRA banks.  SNOMAN processors  unpack the data and
then calibrate it by applying pedestal, slope and offsets to the ADC
data to produce PMT hits with corrected times and charges (other
processors exist to uncalibrate and then pack the Monte Carlo data,
thus allowing the whole data stream to be tested with Monte Carlo
data).  After the calibration processor, the data handling is
identical for Monte Carlo and detector data.  A number of processors
then exist to derive additional information from the PMT hit times and
positions.  The first type of these are called ``fitters.''  These
derive an event position and direction from the PMT data.  A number of
different algorithms exist for this purpose.  In addition to the
fitters there are processors called classifiers that derive
information from the angular distribution of hits in the event.  This
information has been shown to be a powerful tool for discriminating
between neutral-current and charged-current events~\cite{snoman:brice}
and for measuring the most important background events
directly~\cite{snoman:chen}.

Once all of these processors have finished the resulting data can be
viewed using two tools--the event viewer and the event analyzer.  The
event viewer is a graphical display of the SNO detector which displays
the various detector elements and PMT hits in 3D and a number of 2D
formats.  The analyzer is a data-driven tool for sifting through the
entries in the data structure (or through quantities derived from the
entries in the data structure by one of a large number of supplied
functions) to find those that satisfy either individual cuts or
logical combinations of a number of cuts applied to different
quantities.  Any entries or quantities which pass the cuts are written
as an entry in a CERNLIB HBOOK ntuple for later additional analysis with
PAW~\cite{snoman:PAW}.  A slightly unusual feature of SNO (from the
point of view of a typical high-energy physics experiment) is that
some events are correlated in time or position (the spallation
products following a through muon, for example).  This information can
be extracted from the data using the time correlation processor, which
allows the analyzer to combine information from different events that
satisfy conditions placed on their relative time or position into a
single ntuple entry.

\section{Conclusion}
\label{sec:Conclusion}

The Sudbury Neutrino Observatory is a large, second-generation
neutrino detector with active fiducial volumes consisting of 1000
tonnes of highly purified heavy water plus about 1000 tonnes of
purified light water.  The detector will provide high statistics data
from natural neutrino sources, with an emphasis on the study of solar
neutrinos from $^8$B decay and whether or not they undergo flavor
oscillations as they propagate from the core of the sun to the earth.
The ability of SNO to distinguish charged-current and neutral-current
neutrino interactions is unique among present-day solar neutrino
detectors that are operating or under construction.  SNO is also the
only neutrino detector that can exclusively identify antineutrino
interactions.  The unusually deep location of SNO and particular
attention to minimization of radioactive backgrounds makes possible
high signal-to-background measurements.

A series of engineering runs to test detector performance and
commission various detector systems began without water present in the
detector cavity in September, 1997.  SNO was completely filled with
heavy and light water on April 30, 1999, at which time full system
operation commenced.  The detector will be run with pure $\DO$ until
the SNO collaboration determines that a physically interesting dataset
has been accumulated.  Then, the detector will be modified to enhance
its neutral-current detection capabilities.

The resulting dataset should enable SNO to make a definitive
measurement of the fraction of electron neutrinos in the solar
neutrino flux, distinct from all previous measurements in that it will
be independent of theoretical models of the sun.  This and other
measurements are expected to deepen the understanding of the
properties of neutrinos and of the astrophysical sites from which they
originate.


\ack{
This research has been financially supported in Canada by the Natural
Sciences and Engineering Research Council, Industry Canada, National
Research Council of Canada, Northern Ontario Heritage Fund Corporation
and the Province of Ontario, in the United States by the Department of
Energy, and in the United Kingdom by the Science and Engineering
Research Council and the Particle Physics and Astronomy Research
Council.  Further support was provided by INCO, Atomic Energy of
Canada Limited (AECL), Agra-Monenco, Canatom, Canadian
Microelectronics Corporation and Northern Telecom.  The heavy water
has been loaned by AECL with the cooperation of Ontario Hydro.  The
provision of INCO of an underground site is greatly appreciated.  The
collaboration wishes to express gratitude for all the support provided
to make the experiment possible.}

\end{document}